\expandafter\ifx\csname amssym.def\endcsname\relax \else\endinput\fi
\expandafter\edef\csname amssym.def\endcsname{%
       \catcode`\noexpand\@=\the\catcode`\@\space}
\catcode`\@=11
\def\undefine#1{\let#1\undefined}
\def\newsymbol#1#2#3#4#5{\let\next@\relax
 \ifnum#2=\@ne\let\next@\msafam@\else
 \ifnum#2=\tw@\let\next@\msbfam@\fi\fi
 \mathchardef#1="#3\next@#4#5}
\def\mathhexbox@#1#2#3{\relax
 \ifmmode\mathpalette{}{\m@th\mathchar"#1#2#3}%
 \else\leavevmode\hbox{$\m@th\mathchar"#1#2#3$}\fi}
\def\hexnumber@#1{\ifcase#1 0\or 1\or 2\or 3\or 4\or 5\or 6\or 7\or 8\or
 9\or A\or B\or C\or D\or E\or F\fi}

\font\tenmsa=msam10
\font\sevenmsa=msam7
\font\fivemsa=msam5
\newfam\msafam
\textfont\msafam=\tenmsa
\scriptfont\msafam=\sevenmsa
\scriptscriptfont\msafam=\fivemsa
\edef\msafam@{\hexnumber@\msafam}
\mathchardef\dabar@"0\msafam@39
\def\dashrightarrow{\mathrel{\dabar@\dabar@\mathchar"0\msafam@4B}}
\def\dashleftarrow{\mathrel{\mathchar"0\msafam@4C\dabar@\dabar@}}

\def\ulcorner{\delimiter"4\msafam@70\msafam@70 }
\def\urcorner{\delimiter"5\msafam@71\msafam@71 }
\def\llcorner{\delimiter"4\msafam@78\msafam@78 }
\def\lrcorner{\delimiter"5\msafam@79\msafam@79 }
\def\yen{{\mathhexbox@\msafam@55 }}
\def\checkmark{{\mathhexbox@\msafam@58 }}
\def\circledR{{\mathhexbox@\msafam@72 }}
\def\maltese{{\mathhexbox@\msafam@7A }}

\font\tenmsb=msbm10
\font\sevenmsb=msbm7
\font\fivemsb=msbm5
\newfam\msbfam
\textfont\msbfam=\tenmsb
\scriptfont\msbfam=\sevenmsb
\scriptscriptfont\msbfam=\fivemsb
\edef\msbfam@{\hexnumber@\msbfam}
\def\Bbb#1{{\fam\msbfam\relax#1}}
\def\widehat#1{\setbox\z@\hbox{$\m@th#1$}%
 \ifdim\wd\z@>\tw@ em\mathaccent"0\msbfam@5B{#1}%
 \else\mathaccent"0362{#1}\fi}
\def\widetilde#1{\setbox\z@\hbox{$\m@th#1$}%
 \ifdim\wd\z@>\tw@ em\mathaccent"0\msbfam@5D{#1}%
 \else\mathaccent"0365{#1}\fi}
\font\teneufm=eufm10
\font\seveneufm=eufm7
\font\fiveeufm=eufm5
\newfam\eufmfam
\textfont\eufmfam=\teneufm
\scriptfont\eufmfam=\seveneufm
\scriptscriptfont\eufmfam=\fiveeufm
\def\frak#1{{\fam\eufmfam\relax#1}}

\csname amssym.def\endcsname

\expandafter\ifx\csname pre amssym.tex at\endcsname\relax \else \endinput\fi
\expandafter\chardef\csname pre amssym.tex at\endcsname=\the\catcode`\@
\catcode`\@=11
\newsymbol\boxdot 1200
\newsymbol\boxplus 1201
\newsymbol\boxtimes 1202
\newsymbol\square 1003
\newsymbol\blacksquare 1004
\newsymbol\centerdot 1205
\newsymbol\lozenge 1006
\newsymbol\blacklozenge 1007
\newsymbol\circlearrowright 1308
\newsymbol\circlearrowleft 1309
\undefine\rightleftharpoons
\newsymbol\rightleftharpoons 130A
\newsymbol\leftrightharpoons 130B
\newsymbol\boxminus 120C
\newsymbol\Vdash 130D
\newsymbol\Vvdash 130E
\newsymbol\vDash 130F
\newsymbol\twoheadrightarrow 1310
\newsymbol\twoheadleftarrow 1311
\newsymbol\leftleftarrows 1312
\newsymbol\rightrightarrows 1313
\newsymbol\upuparrows 1314
\newsymbol\downdownarrows 1315
\newsymbol\upharpoonright 1316
 
\newsymbol\downharpoonright 1317
\newsymbol\upharpoonleft 1318
\newsymbol\downharpoonleft 1319
\newsymbol\rightarrowtail 131A
\newsymbol\leftarrowtail 131B
\newsymbol\leftrightarrows 131C
\newsymbol\rightleftarrows 131D
\newsymbol\Lsh 131E
\newsymbol\Rsh 131F
\newsymbol\rightsquigarrow 1320
\newsymbol\leftrightsquigarrow 1321
\newsymbol\looparrowleft 1322
\newsymbol\looparrowright 1323
\newsymbol\circeq 1324
\newsymbol\succsim 1325
\newsymbol\gtrsim 1326
\newsymbol\gtrapprox 1327
\newsymbol\multimap 1328
\newsymbol\therefore 1329
\newsymbol\because 132A
\newsymbol\doteqdot 132B
 
\newsymbol\triangleq 132C
\newsymbol\precsim 132D
\newsymbol\lesssim 132E
\newsymbol\lessapprox 132F
\newsymbol\eqslantless 1330
\newsymbol\eqslantgtr 1331
\newsymbol\curlyeqprec 1332
\newsymbol\curlyeqsucc 1333
\newsymbol\preccurlyeq 1334
\newsymbol\leqq 1335
\newsymbol\leqslant 1336
\newsymbol\lessgtr 1337
\newsymbol\backprime 1038
\newsymbol\risingdotseq 133A
\newsymbol\fallingdotseq 133B
\newsymbol\succcurlyeq 133C
\newsymbol\geqq 133D
\newsymbol\geqslant 133E
\newsymbol\gtrless 133F
\newsymbol\sqsubset 1340
\newsymbol\sqsupset 1341
\newsymbol\vartriangleright 1342
\newsymbol\vartriangleleft 1343
\newsymbol\trianglerighteq 1344
\newsymbol\trianglelefteq 1345
\newsymbol\bigstar 1046
\newsymbol\between 1347
\newsymbol\blacktriangledown 1048
\newsymbol\blacktriangleright 1349
\newsymbol\blacktriangleleft 134A
\newsymbol\vartriangle 134D
\newsymbol\blacktriangle 104E
\newsymbol\triangledown 104F
\newsymbol\eqcirc 1350
\newsymbol\lesseqgtr 1351
\newsymbol\gtreqless 1352
\newsymbol\lesseqqgtr 1353
\newsymbol\gtreqqless 1354
\newsymbol\Rrightarrow 1356
\newsymbol\Lleftarrow 1357
\newsymbol\veebar 1259
\newsymbol\barwedge 125A
\newsymbol\doublebarwedge 125B
\undefine\angle
\newsymbol\angle 105C
\newsymbol\measuredangle 105D
\newsymbol\sphericalangle 105E
\newsymbol\varpropto 135F
\newsymbol\smallsmile 1360
\newsymbol\smallfrown 1361
\newsymbol\Subset 1362
\newsymbol\Supset 1363
\newsymbol\Cup 1264
 
\newsymbol\Cap 1265
 
\newsymbol\curlywedge 1266
\newsymbol\curlyvee 1267
\newsymbol\leftthreetimes 1268
\newsymbol\rightthreetimes 1269
\newsymbol\subseteqq 136A
\newsymbol\supseteqq 136B
\newsymbol\bumpeq 136C
\newsymbol\Bumpeq 136D
\newsymbol\lll 136E
 
\newsymbol\ggg 136F
 
\newsymbol\circledS 1073
\newsymbol\pitchfork 1374
\newsymbol\dotplus 1275
\newsymbol\backsim 1376
\newsymbol\backsimeq 1377
\newsymbol\complement 107B
\newsymbol\intercal 127C
\newsymbol\circledcirc 127D
\newsymbol\circledast 127E
\newsymbol\circleddash 127F
\newsymbol\lvertneqq 2300
\newsymbol\gvertneqq 2301
\newsymbol\nleq 2302
\newsymbol\ngeq 2303
\newsymbol\nless 2304
\newsymbol\ngtr 2305
\newsymbol\nprec 2306
\newsymbol\nsucc 2307
\newsymbol\lneqq 2308
\newsymbol\gneqq 2309
\newsymbol\nleqslant 230A
\newsymbol\ngeqslant 230B
\newsymbol\lneq 230C
\newsymbol\gneq 230D
\newsymbol\npreceq 230E
\newsymbol\nsucceq 230F
\newsymbol\precnsim 2310
\newsymbol\succnsim 2311
\newsymbol\lnsim 2312
\newsymbol\gnsim 2313
\newsymbol\nleqq 2314
\newsymbol\ngeqq 2315
\newsymbol\precneqq 2316
\newsymbol\succneqq 2317
\newsymbol\precnapprox 2318
\newsymbol\succnapprox 2319
\newsymbol\lnapprox 231A
\newsymbol\gnapprox 231B
\newsymbol\nsim 231C
\newsymbol\ncong 231D
\newsymbol\diagup 231E
\newsymbol\diagdown 231F
\newsymbol\varsubsetneq 2320
\newsymbol\varsupsetneq 2321
\newsymbol\nsubseteqq 2322
\newsymbol\nsupseteqq 2323
\newsymbol\subsetneqq 2324
\newsymbol\supsetneqq 2325
\newsymbol\varsubsetneqq 2326
\newsymbol\varsupsetneqq 2327
\newsymbol\subsetneq 2328
\newsymbol\supsetneq 2329
\newsymbol\nsubseteq 232A
\newsymbol\nsupseteq 232B
\newsymbol\nparallel 232C
\newsymbol\nmid 232D
\newsymbol\nshortmid 232E
\newsymbol\nshortparallel 232F
\newsymbol\nvdash 2330
\newsymbol\nVdash 2331
\newsymbol\nvDash 2332
\newsymbol\nVDash 2333
\newsymbol\ntrianglerighteq 2334
\newsymbol\ntrianglelefteq 2335
\newsymbol\ntriangleleft 2336
\newsymbol\ntriangleright 2337
\newsymbol\nleftarrow 2338
\newsymbol\nrightarrow 2339
\newsymbol\nLeftarrow 233A
\newsymbol\nRightarrow 233B
\newsymbol\nLeftrightarrow 233C
\newsymbol\nleftrightarrow 233D
\newsymbol\divideontimes 223E
\newsymbol\varnothing 203F
\newsymbol\nexists 2040
\newsymbol\Finv 2060
\newsymbol\Game 2061
\newsymbol\mho 2066
\newsymbol\eth 2067
\newsymbol\eqsim 2368
\newsymbol\beth 2069
\newsymbol\gimel 206A
\newsymbol\daleth 206B
\newsymbol\lessdot 236C
\newsymbol\gtrdot 236D
\newsymbol\ltimes 226E
\newsymbol\rtimes 226F
\newsymbol\shortmid 2370
\newsymbol\shortparallel 2371
\newsymbol\smallsetminus 2272
\newsymbol\thicksim 2373
\newsymbol\thickapprox 2374
\newsymbol\approxeq 2375
\newsymbol\succapprox 2376
\newsymbol\precapprox 2377
\newsymbol\curvearrowleft 2378
\newsymbol\curvearrowright 2379
\newsymbol\digamma 207A
\newsymbol\varkappa 207B
\newsymbol\Bbbk 207C
\newsymbol\hslash 207D
\undefine\hbar
\newsymbol\hbar 207E
\newsymbol\backepsilon 237F
\catcode`\@=\csname pre amssym.tex at\endcsname

\font\teneusm=eusm10
\newfam\eusmfam
\textfont\eusmfam=\teneusm
\def\scr#1{{\fam\eusmfam\relax#1}}
\def\ad{\mbox{{\mathrm ad\thinspace}}}

\def\l{\left}
\def\r{\right}
\def\la{\langle}
\def\ra{\rangle}
\documentstyle{article}
\begin{document}
\newtheorem{theorem}{Theorem}[section]
\newtheorem{corollary}[theorem]{Corollary}
\newtheorem{lemma}[theorem]{Lemma}
\newtheorem{proposition}[theorem]{Proposition}
\newtheorem{remark}[theorem]{Remark}
\author{Elizabeth Jurisich$^*$ \\ James Lepowsky\\ Robert L. Wilson}
\title{Realizations of the Monster Lie Algebra}
\date{}
\maketitle
\begin{center}
Department of Mathematics, Rutgers University\\
New Brunswick, NJ 08903\\
$^*$Current address: Department of Mathematics, University of Chicago\\
Chicago, IL 60637\\
\end{center}
e-mail addresses: jurisich@math.rutgers.edu, lepowsky@math.rutgers.edu,\\
rwilson@math.rutgers.edu

\section{Introduction}

In this paper we reinterpret the theory of generalized
Kac-Moody Lie algebras in terms of local Lie algebras formed from
reductive or Kac-Moody algebras and certain modules (possibly
infinite-dimensional) for these algebras. We exploit the fact,
established in \cite{Jur2}, that
certain generalized Kac-Moody algebras contain specific large
free subalgebras, in order to exhibit these generalized Kac-Moody algebras
as explicitly prescribed Lie algebras of operators acting on tensor
algebras. In the most important special case, that of R. Borcherds'
Monster Lie algebra $\frak m$, introduced in \cite{B3} (see also
\cite{B4}), we apply the
free subalgebra result \cite{Jur2} to simplify Borcherds' work \cite{B3}
on the Conway-Norton conjectures (see \cite{CN}) for the ``moonshine
module'' $V^\natural$ (\cite{FLM1}, \cite{FLM}) for the Fischer-Griess
Monster group $M$. In particular, we realize $\frak m$ as an
explicitly prescribed $M$-covariant Lie
algebra of operators acting on the tensor algebra over a certain
$\frak g \frak l_2$- and $M$-module built in a simple way from the
$M$-module $V^\natural$.

In \cite{B3}, Borcherds produces recursion relations which, along with
initial conditions, uniquely characterize
the McKay-Thompson series (i.e., the graded
traces) of the elements of $M$ acting on the
infinite-dimensional $M$-module $V^\natural$ constructed in
\cite{FLM1}, \cite{FLM}. In
this way he completes the proof of the Conway-Norton conjectures
as they relate to $V^\natural$, in the sense that he shows
that the McKay-Thompson series for $V^\natural$ do indeed agree with
the modular
functions listed in \cite{CN} for all elements of $M$, since the
coefficients of those modular functions satisfy the same recursion
relations (replication formulas) and initial conditions. For instance,
for the identity element of $M$ the corresponding McKay-Thompson series is
the modular function $J(q) = j(q) -744$, which is the graded dimension
of $V^\natural$. (This is one of the McKay-Thompson series for $V^\natural$
already determined in \cite{FLM1}, \cite{FLM}.) The
proof in \cite{B3} involves
defining the Lie algebra $\frak m$, establishing a triangular
decomposition $\frak m =\frak n^+ \oplus \frak h \oplus \frak n^-$ and
a denominator identity for $\frak m$, and
computing the homology of the subalgebra $\frak n^+ \subset \frak m$
(or equivalently, of the subalgebra $\frak n^-$).
Borcherds is then able to show that the
coefficients of the McKay-Thompson series satisfy certain recursion relations.
As we will explain below, much of the
theory involved in the proof in \cite{B3} is illuminated
if we decompose $\frak m$ as
$$\frak m= \frak u^+ \oplus \frak g \frak l_2  \oplus \frak u^-,$$
where $ \frak u^+$ and $\frak u^-$ are the Lie algebras proved to be free
in \cite{Jur2}, and if we consider the $\frak g \frak l_2$-module
$\frak u^-$ in place of $\frak n^-$. We base our considerations on
this decomposition
rather than the triangular decomposition used in \cite{B3}.

The properties of the Monster Lie algebra are of central importance to
the proof in \cite{B3}. In Section 2 we recall facts about
generalized Kac-Moody algebras, of which $\frak m$ is an example.
Borcherds' original work on the subject \cite{B1} contains results
such as character formulas and a denominator identity
for generalized Kac-Moody algebras.
V. Kac in \cite{K2} gives an outline (without
detail) of how to rigorously develop the theory of generalized
Kac-Moody algebras by indicating that one should follow the arguments
presented there for Kac-Moody algebras; see also \cite{HMY}. Included in
\cite{Jur1} is an exposition developing the theory of generalized
Kac-Moody algebras, including an extension of the homology results of
\cite{GL}
(not covered in \cite{K2}) to these more general Lie algebras. That this can
be done is mentioned and used in \cite{B3}. This homology theory
gives another proof of the character and denominator formulas (see
\cite{Jur1}). We find it appropriate to work with
the extended Lie algebra as in \cite{GL} and \cite{L1} (that is, the Lie
algebra with suitable degree derivations adjoined). Equivalently, one
can generalize the theorems in \cite{K2}. In either of these
approaches the Cartan subalgebra is sufficiently enlarged to make the
simple roots linearly independent and of multiplicity one.

It is shown in \cite{Jur2} that any generalized Kac-Moody algebra
$\frak g$ that has no mutually orthogonal imaginary simple roots
can be decomposed as
$\frak g = \frak u^+ \oplus (\frak g_S +
\frak h) \oplus \frak u^-$ where $\frak g_S$ is a (semisimple or)
Kac-Moody Lie algebra
and $\frak u^+$ and $\frak u^-$ are free Lie algebras over particular
modules for $\frak g_S$. This theorem is recalled in Section 2 below,
where we also summarize properties of $\frak m$ which we use. It is
also shown that there are Lie algebras which cover $\frak m$ such that
the natural action of $M$ on $\frak m$ lifts to an action of $M$
on these larger generalized Kac-Moody algebras.

Facts about local Lie algebras that we use are also included in
Section 2. Mainly, we recall that the
maximal Lie algebra $\frak a_{max}$ associated to a local Lie algebra
$\frak a=\frak a_1 \oplus \frak a_0 \oplus \frak a_{-1}$ is such that
$\frak a_{max}^\pm$ is the free Lie algebra on $\frak a_{\pm 1}$, as in
\cite{K1}.
In
light of the results of \cite{Jur2} mentioned above, it is not
surprising that these generalized Kac-Moody algebras with
suitable derivations adjoined are
closely related to maximal Lie algebras $\frak a_{max}$ associated to
the appropriate local Lie algebras $\frak a$. We show this in Section 3.

In Section 4, we discuss some examples of standard (irreducible)
modules which have the special property that they are
induced from irreducible modules of the subalgebra $(\frak g_S +
\frak h) \oplus \frak u^+$. As
mentioned above, combined with the free subalgebra result, this
gives yet another natural realization of the Monster
Lie algebra, this time as an explicitly prescribed $M$-covariant Lie
algebra of operators on the space
$T({\scr V})$, where ${\scr V}$ is one of the two
(infinite-dimensional) $\frak g \frak l_2$- and $M$-modules
$\frak a_{\pm 1}$ occurring
in the local Lie algebra construction (here $\frak a_0 = \frak g
\frak l_2$). The examples of standard
modules that we consider in this paper are all ``generalized Verma
modules,'' that is, they are induced from irreducible modules for
``parabolic subalgebras'' (e.g., the subalgebra
$\frak p = (\frak g_S + \frak h) \oplus \frak u^+$ for a generalized
Kac-Moody algebra $\frak g =\frak u^- \oplus (\frak g_S +
\frak h)\oplus \frak u^+$).
The considerations on generalized Verma modules in \cite{GL}
and \cite{L2} remain
applicable to the case of generalized Kac-Moody algebras.
If we choose
the right parabolic subalgebra and highest weight $\lambda$,
then the corresponding generalized Verma module is a
standard module, and in particular, it is irreducible. In the case where
$\frak u^-$ is a free Lie algebra over a vector space $V$, the tensor
algebra over $V$ clearly has the structure of
a generalized Verma module (see Section 4). As we have mentioned, this
includes the case of the Monster Lie algebra
$\frak m$. The ``fundamental modules'' constructed in
\cite{GT} for those generalized Kac-Moody algebras (or Borcherds
algebras, as
they are called in \cite{GT}) with one simple imaginary root are
special cases of these induced modules. For the case of $\frak m$, we
conjecture that the weight-zero components $v_0$ of the vertex
operators $Y(v,z)$ for $v \in \scr V$
generate a free associative algebra of operators on $\frak m$.
This may provide a more
conceptual framework for studying the irreducible $\frak m$-modules
$T(\scr V)$. (See Sections 2 and 4.)

Finally, in Section 5, we use the free Lie algebra
structure of $\frak u^-$ to simplify part of the proof appearing in
\cite{B3} concerning the replication formulas.
We follow the general ideas of Borcherds' proof, but we use the
subalgebra $\frak u^-$ in place of $\frak n^-$. Because of its
structure as a free Lie algebra, computations involving $\frak u^-$ are
comparatively simple. For example, the homology of $\frak u^-$ with
coefficients in the trivial module is elementary to compute, while
providing us with essentially the same information as the homology of
$\frak n^-$.
Moreover, the formula, obtained from the Euler-Poincar\'{e} identity,
that leads to replication (now symmetric in the
formal variables $p$
and $q$) is visibly similar to formulas occurring in such classical
references on free Lie algebras as \cite{Bou} and \cite{Serre} (as is
the proof of the
denominator identities in \cite{Jur2}). We are able to obtain
the replication formula as it appears in \cite{No} by
carrying out an analog of a computation in \cite{Bou}. After proving a
M\"{o}bius inversion formula consistent with Adams operations, we establish
recursion relations equivalent to replication.
Even though replication is more
easily understood in this context, it is still not clear why series
satisfying these replication formulas should automatically have
modular transformation properties.
Philosophically, the fact that $\frak n^+$ (or $\frak n^-$)
contains such a large free Lie algebra explains why computations such
as determining root multiplicities and the replication formulas are
at all manageable for $\frak m$.

This paper was presented by E. J. and J. L. at the June, 1994
AMS-IMS-SIAM Joint Summer Research Conference on Moonshine, the
Monster and Related Topics at Mount Holyoke College.

We are grateful to S.-J. Kang for informing us that formula (37) in
this paper has been independently obtained in joint work with V. Kac.

We would like to thank R. Borcherds, F. Knop and C. Weibel for
helpful comments. We also thank R. Gebert and J. Teschner, K.
Harada, M. Miyamoto and H. Yamada, S. J. Kang and S. Naito for sending us
preprints of their works on generalized Kac-Moody algebras.
E. J. is supported in part by a Rutgers
University Louis Bevier Graduate Fellowship, and J. L. and R.~W. are
supported in part by NSF grant DMS-9111945.

\section{Preliminary Material}

\subsection{Notation}

For any vector space $V$ we let $L(V)$ denote the free Lie algebra over
$V$, and $T(V)$ the tensor algebra over $V$. Recall
that $T(V)$ is the universal enveloping algebra of $L(V)$.
Let $V^*$ denote the dual space of $V$.  For a ${\Bbb
Z}$-graded vector space $V = \coprod_{i \in \Bbb Z } V_i$ we let
$V^{\pm}$ denote $\coprod_{i \in {\pm} \Bbb Z_+}V_i$.
Unless otherwise indicated, vector spaces will be over $\Bbb C$.
We denote the Monster group by $M$ and the
``moonshine module'' for $M$ by $V^\natural$, as in
\cite{FLM}.

\subsection{Generalized Kac-Moody algebras}

The theory of generalized Kac-Moody Lie algebras with imaginary simple
roots was developed by
Borcherds starting in \cite{B1}. Two further
definitions (different from the first and from one another) are
given in \cite{B2} and \cite{B3}. Of these definitions
the one given below most resembles the one appearing in \cite{B1},
where there is a definition in terms of an index set $I$, a vector
space $H$, a set of generators $H\cup \{e_i,f_i| \ i \in I\}$, and
relations, determined by square matrices, which agree with the ones
given here. Borcherds does not divide out by the radical (that is, the
largest graded ideal having trivial intersection with $\frak h$) in his
definition of generalized Kac-Moody algebra.
In \cite{B2}
Borcherds extends his definition to include some additional
elements $h_{ij}$ which are central. These elements do not enter into
our discussion.
In \cite{B3} Borcherds defines generalized
Kac-Moody algebras in terms of ``an almost positive definite bilinear
form'' and an involution, and it is a theorem stated in \cite{B3} (see also
\cite{HMY} and \cite{Jur2} for clarification of the statement and for
two different detailed proofs; cf. \cite{K2}) that such
Lie algebras are closely related to Lie algebras defined from matrices.

The definition given here of the generalized Kac-Moody algebra
$\frak g(A)$ (see below) is the same as that in
\cite{Jur2}. In \cite{Jur1} and \cite{K2} a generalized
Kac-Moody algebra is defined as $\frak g(A)$ modulo
its radical. The radical of
$\frak g(A)$ is in fact zero (see
\cite{HMY} and \cite{Jur1} for detailed proofs, clarifying \cite{B1}, using
\cite{GK} or \cite{K2}; or see \cite{Jur2}, where there is a relatively
elementary argument for the case of Lie algebras satisfying the
hypothesis of Theorem~\ref{thm:free} below---which are the main
examples discussed in this paper---showing that the vanishing of the
radical for Kac-Moody algebras implies the vanishing of the radical
for this class). Thus the definition given here is
equivalent to those in \cite{Jur1} and \cite{K2}.

Theorem~\ref{thm:free}, given at the end of this section, is the main
theorem of \cite{Jur2}, and is used throughout this paper.

We will use the notation
of \cite{GL}, \cite{L1} and \cite{L2}, suitably generalized
to the case
of generalized Kac-Moody Lie algebras in \cite{Jur1}. We restrict our
attention to the case of symmetric matrices (although symmetrizable
matrices could be easily handled) and countable index sets.

Let $I$ be a (finite or) countable index set and let $A = (a_{ij})_{i,j \in
I}$ be a
matrix with entries in ${\Bbb R}$, satisfying the following conditions:
\begin{description}
\item[(C1)]  $A$ is symmetric.
\item[(C2)]  If $ i\neq j$ ($i,j \in I$), then $a_{ij}~\leq~0 $.
\item[(C3)]  If $a_{ii} > 0$ ($i \in I$), then ${2a_{ij} \over a_{ii}}
\in \Bbb Z $ for all $j  \in I$.
\end{description}
Let $\frak g(A)$ be the Lie algebra with generators
$h_{i}, e_i, f_i$, $i \in I$, and the following defining
relations: For all $i,j,k \in I$,

\begin{description}
 \item[(R1)] $\left[ h_{i}, h_{j}\right] =0$,
 \item[(R2)] $\left[ h_{i}, e_k\right]   -   a_{ik} e_k =0$,
 \item[(R3)] $\l[h_{i}, f_k\r]  +   a_{ik} f_k =0$,
 \item[(R4)] $\l[e_i , f_j \r] -   \delta_{ij} h_{i} =0$,
 \item[(R5)] $ (\ad e_i)^{{-2a_{ij} \over a_{ii}} + 1}e_j =0$ and
        $(\ad f_i)^{{-2a_{ij} \over a_{ii}} + 1}f_j =0
\mbox{ for all } i \neq j \mbox{ with } a_{ii} > 0$,
 \item[(R6)] $[e_i, e_j]=0$ and $ [f_i,f_j]=0$ whenever $a_{ij} =0$.
\end{description}

\noindent {\bf Definition}: The Lie algebra $\frak g(A)$ is {\em the
generalized Kac-Moody (Lie) algebra associated to the matrix A}.
Any
Lie algebra of the form $\frak g(A) / \frak c$ where $\frak c$ is a
central ideal is called a {\em generalized Kac-Moody algebra}. \vspace{1ex}

In $\frak g(A)$, the elements $h_i,e_i,f_i$ for $i \in I$ are linearly
independent. Define the {\em Cartan subalgebra} $\frak h$ of $\frak
g(A)$ to be the span of
the $h_i$ for ${i\in I}$. The center of $\frak g(A)$ lies in $\frak h$.

Let $A$ satisfy $(C1)- (C3)$ and let $S \subset I$ be the set of all
indices $i$ such that $a_{ii} > 0$. (Many of the following considerations
apply for more general subsets of $I$; it is for convenience that we
fix $S$ as indicated.) Let $B$ be the
the submatrix  of $A$ given by $(a_{ij})_{i,j \in S}$.

The associated Lie algebra
$\frak g(B)$ is a Kac-Moody algebra (with a symmetric defining matrix).
Let  $\frak g_S$ be the Lie
subalgebra of $\frak g(A)$ generated by $\{e_i, h_i,f_i\}_{i\in S}$.
Then $\frak g_S$ is isomorphic to $\frak g(B)$. Note that we are using
the characterization of $\frak g(B)$ in terms of
generators and relations, i.e., we are using the fact that the radical
is zero (see \cite{GK} or \cite{K2}; this is Serre's theorem \cite{S2}
in case $B$ is of finite type).
For the important case of the Monster Lie algebra $\frak m$, described
in more detail in the next section, $B =(2)$
and $\frak g_S \cong \frak s \frak l_2$.

The Lie algebra $\frak g = \frak g(A)$ is naturally graded by ${\Bbb Z}^I$.
Let $D_i$ be the $i^{th}$ degree derivation of $\frak g$ with respect
to this
grading, i.e., $D_i e_j = \delta_{ij}e_j$, $D_i f_j = -\delta_{ij}f_j$
and $D_i h_j =0$ for all $i,j \in I$. The span of the $D_i$ for $i \in
I$ forms an abelian Lie
algebra $\frak d_0$ of derivations of $\frak g$. Given a subspace
$\frak d$ of $\frak d_0$ let $\frak g^e$ be the semidirect product $\frak d
\ltimes \frak g$. Denote by $\frak h^e$ the abelian subalgebra $\frak d
\oplus \frak h$. The Lie algebra $\frak g^e$ is called the {\em extended} Lie
algebra. Call $\frak h^e$ the {\em Cartan subalgebra} of $\frak g^e$.

Define the {\em simple roots} $\alpha_i \in
(\frak h^e)^*$ for $i \in I$ by the conditions
$[h,e_i]=\alpha_i(h)e_i$ for all $h \in (\frak h^e)^*$, $i \in I$.
Fix a subspace $\frak d$ of $\frak d_0$ such that
the simple roots $\alpha_i$ are linearly independent. There is a
unique symmetric bilinear form $(\cdot,\cdot)$ on the span of the
$\alpha_i$, $i \in I$, determined by the conditions $(\alpha_i,\alpha_j)
=a_{ij}$.
This can be
extended (not necessarily uniquely) to a symmetric bilinear form, which we
continue to call $(\cdot,\cdot)$, on all of $(\frak h^e)^*$ such that
$(\varphi,\alpha_i) = \varphi (h_i)$ for all $\varphi \in (\frak h^e)^*$.

For $\alpha \in (\frak h^e)^*$ define
$$\frak g^\alpha = \{x \in \frak g | [h,x] = \alpha (h)x \mbox{ for
all } h \in \frak h^e\}.$$ Nonzero elements $\alpha \in (\frak h^e)^*$
such that
$\frak g^\alpha \neq 0$ are called {\em roots}.
Let $\Delta \subset (\frak h^e)^*$ be the set of roots, and $\Delta_+$ the
set of roots which are
nonnegative integral linear combinations of $\alpha_i$'s.
The roots in $\Delta_+$ are called {\it positive} roots. Let $\Delta_- =
-\Delta _+ $,
the set of {\it negative} roots. All of the roots are either
positive or negative.
There is a root space decomposition
\begin{equation}
\frak g = \coprod_{\varphi \in \Delta_+} \frak g^{\varphi} \oplus
\frak h \oplus \coprod_{\varphi \in \Delta_-}\frak g^{\varphi}.
\label{eq:rts}\end{equation}

Define the set of {\em real} roots $\Delta_R$ to be the set of $\alpha
\in \Delta$ such that $(\alpha, \alpha)>0$. The set of {\em
imaginary} roots is $\Delta \backslash \Delta_R$. The imaginary simple
roots are those $\alpha_i$ for which $a_{ii} \leq 0$.

The symmetric bilinear form on the span of the $\alpha_i$, $i \in I$,
can be transferred to $\frak h$ by letting $(h_i,h_j)_{\frak h} =
(\alpha_i, \alpha_j) = a_{ij}$. This form can be extended (not
uniquely, for it depends upon a choice of symmetric bilinear form on
$\frak d$) to a
symmetric bilinear form $(\cdot, \cdot)_{\frak h^e}$ on $\frak h^e= \frak d
\oplus \frak h$, satisfying the conditions $(d,h_j)= \alpha_j(d)$ for
$d \in \frak d$ and $j \in I$. We assume that this form on $\frak h^e$
is consistent with the form on $(\frak h^e)^*$. There is a unique invariant
symmetric bilinear form $(\cdot, \cdot)_{\frak g^e}$ on $\frak g^e$
which extends
$(\cdot, \cdot)_{\frak h^e}$. This form satisfies the condition
$$(\frak g^\varphi, \frak g^\psi)_{\frak g^e} =
(\frak h^e,g^\varphi)_{\frak g^e} =0$$ for all $\varphi,
\psi \in \Delta $ such that $\varphi + \psi \neq 0$.
Given any $\varphi = \sum_{i
\in I} n_i \alpha_i \in \Delta$ ($n_i \in \Bbb Z$), we define
$h_{\varphi} = \sum_{i\in I}n_i h_i$. Then for $a \in \frak g^\varphi$,
$b \in \frak g^{-\varphi}$, the
form $(\cdot,\cdot)_{\frak g^e}$ satisfies the condition
$$[a,b]=(a,b)h_{\varphi}.$$
In particular, $(e_i,f_j)_{\frak g^e}= \delta_{ij}$ for $i,j \in I$. (See
\cite{Jur1} for details; cf. \cite{K1}, \cite{L1}, \cite{Mo}.)

We assume that in the case in which $A$ is a (finite) generalized
Cartan matrix, the space $\frak d$ and the form on $\frak d$ are chosen
so that the form $(\cdot, \cdot)_{\frak h^e}$ on the
(finite-dimensional) space $\frak h^e$ is nonsingular, so that the Lie
algebra $\frak g^e=\frak g(A)^e$ is, in this case,  the usual
Kac-Moody algebra (cf. \cite{K2}) and the form
$(\cdot,\cdot)_{\frak g^e}$ is nondegenerate.

The center of the unextended Lie algebra $\frak g$ is equal to
the radical of the form $(\cdot,\cdot)$ restricted to $\frak g$. We
also note that any
element of the center $h =\sum_{j \in I} c_{j}h_{j}$ ($c_{j} \in \Bbb
C$,  $c_j=0$ for almost all $j \in I$)
provides a linear dependence relation $\sum_{j \in I}c_{j}a_{ij}$,
among the columns of the matrix $A$ and conversely, any such
relation among columns of $A$ determines an element of the center of
$\frak g$.

We make the following definitions:
$\Delta^S = \Delta \cap \coprod_{i \in S} {\Bbb Z} \alpha_i$,
$\Delta^S_+ = \Delta_+ \cap \Delta^S $ and $\Delta^S_- = \Delta_-
\cap \Delta^S$. Denote by $\frak h_S$ the span of the $h_i$, ${i \in
S}$. Then we have the root space decomposition:
\begin{equation}
\frak g_S =\coprod_{\varphi \in \Delta^S_+} \frak g^{\varphi} \oplus
\frak h_S \oplus \coprod_{\varphi \in \Delta^S_-}\frak g^{\varphi}.
\label{eq:root} \end{equation}
Define the following subalgebras of $\frak g = \frak g(A)$:
$$
\frak n^+ = \coprod_{\varphi \in \Delta_+}\frak g^{\varphi};\quad
\frak n^- = \coprod_{\varphi \in \Delta_-}\frak g^{\varphi};\quad
\frak n^+_S = \coprod_{\varphi \in \Delta^S_+}\frak g^{\varphi};
$$
\begin{equation}
\frak n^-_S = \coprod_{\varphi \in \Delta^S_-}\frak g^{\varphi};\quad
\frak u^+ =\coprod_{\varphi \in \Delta_+ \backslash \Delta^S_+}
                  \frak g^{\varphi};\quad
\frak u^-= \coprod_{\varphi \in \Delta_- \backslash \Delta^S_-}
                  \frak g^{\varphi};\quad
\frak r =\frak g_S + \frak h .\label{eq:dec}
\end{equation}

Let $\frak p = \frak r \oplus \frak u^+$, the ``parabolic subalgebra'' of
$\frak g$ defined by $S$. Note that
$$
\frak g = \frak u^- \oplus \frak p.
$$
The subalgebras $\frak p^e$, $\frak r^e$ and $\frak g^e_S$
are defined in the obvious ways.

Let $X$ be an $\frak h^e$-module, e.g., a $\frak
g^e$-module considered as an $\frak h^e$-module by restriction, and
let  $\nu \in
(\frak h^e)^*$. Define the corresponding {\it weight space}
$$X_{\nu}=\{x \in X | h \cdot x = \nu (h)x \mbox{ for all } h \in
\frak h^e\}.$$
We say that $\nu$ is a {\it weight } if $X_\nu \neq 0$ and that the
nonzero elements of $X_{\nu}$ are {\em weight vectors} of weight $\nu$.
The module $X$ is called a {\it weight module} if $X$ is a direct sum
of its weight spaces.

 A $\frak g^e$-module is called a {\it highest weight module} if it is
generated by a weight vector $x$ annihilated by $\frak n^+$. The
weight of $x$ is called the {\em highest weight} of $X$.
For any highest weight
module $X$ we have ${U}(\frak n^-) \cdot x =X$. Lowest weight
modules are defined analogously.

Concepts defined above for $\frak g(A)$ such as extended Lie
algebra, roots, simple
roots, Cartan subalgebra and weights can be defined analogously
for a generalized Kac-Moody algebra of the form $\frak g(A)/ \frak c$
with $\frak c$ a subspace of the center.
Note that in particular, the roots of $\frak g(A)$, which are elements
of $(\frak h^e)^* $,  vanish on $\frak c \subset
\frak h$ and may therefore be identified with the roots of
$\frak g(A)/\frak c$, which are elements of the dual of
$(\frak h/\frak c)^e = \frak d \oplus \frak h/\frak c$.
We shall thereby identify
the roots, simple roots, the $e_i$'s and the $f_i$'s (but not the
$h_i$'s), etc., of $\frak g(A)$ with those of $\frak g(A)
/ \frak c$. The simple roots remain linearly independent. For those
$\frak g(A)^e$ weight modules $X$ on which
$\frak c$ acts trivially, $X$ is also a $(\frak g(A)/ \frak c)^e$-module,
and we identify the weights, etc., for $X$ as $\frak g(A)^e$
and $(\frak g(A)/ \frak c)^e$-modules.

We shall use the following theorem from \cite{Jur2}:

\begin{theorem}
Let $\frak g(A)= \frak g$ be a generalized Kac-Moody algebra such that if
$\alpha_i$ and $\alpha_j$ are two distinct imaginary simple roots then
$a_{ij} <0$. Let $S = \{ i \in I | a_{ii}>0\}$.
Then $\frak g = \frak u^+ \oplus \frak r \oplus \frak u^-$, where
$\frak u^-$ is the free Lie algebra
on the direct sum of the integrable highest weight
$\frak g_S$-modules
${U}(\frak n^-_S)\cdot f_j$ for $j \in I\backslash S$ and
$\frak u^+$ is the free Lie algebra on the direct sum of the
integrable lowest weight $\frak g_S$-modules
${U}(\frak n^+_S)\cdot e_j$ for $j \in I\backslash S$.
\label{thm:free}
\end{theorem}

\subsection{The Monster Lie algebra}

The Monster Lie algebra $\frak m$ is defined by Borcherds in \cite{B3}
(see \cite{Jur2} for more detail and further references).
It is constructed using the vertex operator
algebra $V^\natural$ and the vertex algebra
$V_L$ associated to an even Lorentzian lattice $L$ of rank 2, which we
identify with $\Bbb Z \oplus \Bbb Z$, equipped with the bilinear form
$\la \cdot,\cdot \ra$ given by the matrix
$\l(\begin{array}{cc}
0 & -1\\
-1 &0
\end{array}\r)$. The Lie algebra $\frak m$ is
a certain quotient of the ``physical subspace'' $P_1$ of
the vertex algebra $V^\natural \otimes V_L$ ($P_1$ is the space of
lowest weight vectors of weight one for the Virasoro algebra):
$ \frak m =P_1/ R $, where $R$ is the radical of a natural symmetric
bilinear form.
We do not need the
details of the actual definition in
this paper, but will quote results from \cite{B3}, \cite{FLM} and
\cite{Jur2} as they are needed. (We note that Borcherds' construction of
the Lie algebra is over $\Bbb R$, while we are working over $\Bbb C$.)

The vertex operator algebra $V^\natural$ is equipped with an action of
$M$, and $V^\natural$ has a natural $M$-invariant
symmetric nondegenerate bilinear form (see \cite{FLM}).

We grade $V^\natural$ as $\coprod_{i \geq -1}
V^\natural_i$ where $V^\natural_i$ is the span of all vectors of
conformal weight $i+1$. Recall (\cite{FLM1}, \cite{FLM})
that $V^\natural$ has graded dimension $\sum_{i \geq -1}
\l(\dim V^\natural_i\r) q^i =
J(q)$, where $J(q) =\sum_{i \geq -1} c(i)q^i$ is the
modular function $j(q) - 744$, so that $\dim V^\natural_i = c(i)$ for
all $i \geq-1$. We have $c(-1) =1,\ c(0)=0,\ c(1)=196884$. (For $i
<-1$ we set $V^\natural_i =0$ and $c(i)=0$.) The action of $M$
preserves each $V^\natural_i$.

The rank $2$ lattice $L$ provides a $\Bbb Z
\oplus \Bbb Z$-grading of the Lie algebra $\frak m$.
The group action of $M$ on the vertex operator algebra $V^\natural$
gives an action of $M$ on the vertex algebra $V^\natural \otimes V_L$,
where $M$ acts trivially on $V_L$.
This action induces an action of $M$ as Lie algebra automorphisms on
$\frak m$, preserving the grading.
The Lie algebra $\frak m$
has a natural nondegenerate invariant symmetric bilinear form which is
also $M$-invariant. This form satisfies the condition $(\frak m_r,
\frak m_s) = 0$ unless $r+s=0$, for $r,s \in \Bbb Z \oplus \Bbb Z$.
Lemma \ref{lem:ng} below is
one of the main theorems of \cite{B3} (cf. \cite{Jur2} for some
clarification), and follows
{}from the ``no-ghost'' theorem from string theory. Although the
statement of the theorem \cite{GTh} involves vector
spaces over $\Bbb R$, after
the linear isomorphism given below is established over $\Bbb R$ we
extend the isomorphism to $\Bbb C$.

\begin{lemma}
As a $\Bbb Z\oplus \Bbb Z$-graded Lie algebra and $M$-%
module, $\frak m$ satisfies:
\begin{equation}
\frak m = \l(\coprod_{m,n <0}\frak m_{(m,n)} \r)\oplus
\frak m_{(-1,1)}\oplus \frak m_{(0,0)}\oplus  \frak m_{(1,-1)} \oplus
\l(\coprod_{m,n >0}\frak m_{(m,n)}\r),\label{eq:ng}
\end{equation}
where there are natural isomorphisms
$$\frak m_{(m,n)} \cong V^\natural_{mn} \mbox{ as an $M$-module for }
(m,n) \neq (0,0),$$
$$\frak m_{(0,0)} \cong \Bbb C \oplus \Bbb C, \mbox{ a trivial
$M$-module}.$$
Furthermore, the $M$-invariant bilinear form is preserved under these
isomorphisms, i.e., the pairing between $\frak m_{(m,n)}$ and $\frak
m_{(-m,-n) }$ corresponds to the bilinear form on $V^\natural_{mn}$.
$\square$

\label{lem:ng}
\end{lemma}

We define
$$ \frak n^+ = \coprod_{m,n >0} \frak m_{(m,n)} \oplus
\frak m_{(1,-1)},$$
and
$$ \frak n^- = \coprod_{m,n <0} \frak m_{(m,n)} \oplus
\frak m_{(-1,1)},$$
so that
$$\frak m = \frak n^- \oplus \frak m_{(0,0)} \oplus \frak n^+.$$

It follows from the definition of $\frak m$ that
$$\frak m_{(-1,1)} \oplus \frak m_{(0,0)}\oplus  \frak m_{(1,-1)} \cong
\frak g \frak l_2.$$
To make this explicit, let
$$ 1 \rightarrow \la
\kappa\ra \rightarrow \hat L {\buildrel - \over \rightarrow} L
\rightarrow 1$$
be the central extension of $L$ with $\kappa$
of order $2$ and commutator map given by $\kappa^{\la \alpha, \beta\ra}$,
$\alpha , \beta \in L$ (cf. \cite[Ch. 7]{FLM}). We
choose elements $a, b\in \hat L$ to satisfy
$\bar a = (1,1)$, $\bar b = (1,-1) \in L$. As in \cite{Jur2}, appropriate
elements of $V^\natural \otimes V_L$
are used to denote their equivalence classes
in $\frak m$. Let $\iota$ denote the natural map from $\hat L$ to
$V_L$, so that for $c \in  \hat L$, $\iota (c)$ is identified with $ 1
\otimes \iota (c) \in 1 \otimes \Bbb C
\{ L\} \subset S \otimes\Bbb C \{ L\} = V_L $, where $S$ is the
symmetric algebra of the negatively graded subspace of the appropriate
Heisenberg algebra and where $\Bbb C \{ L\}$ is the ``twisted group
algebra'' of $L$ (cf. \cite[Ch. 7]{FLM}).

Consider the elements $e =1\otimes \iota(b)\in \frak m_{(1,-1)},\  f
=1 \otimes \iota(b^{-1})\in
\frak m_{(-1, 1)} ,\  h =\bar b(-1) \otimes \iota(1)\in \frak m_{(0,0)}$ and
$z=\bar a(-1)\otimes\iota(1)\in \frak m_{(0,0)}$ in $\frak m$.
Then $e,f$ and $h$ span a copy of
$\frak s \frak l_2$, with relations
$[e,f] = h$,
$[h,e]= 2e$ and
$[h,f]= -2f.$ Since $z$ commutes with $e, f$ and $h$, the above elements
span a copy of $\frak g\frak l_2 $ in $\frak m$.
Under the adjoint action $\frak m$ is a $\frak g
\frak l_2$-module, and the element $z$ acts
on $\frak m_{(m,n)}$ as the scalar $-m-n$ and $h$ acts on
$\frak m_{(m,n)}$ as the scalar $m-n$. Note that
$\frak m_{(-1,1)} \oplus \frak m_{(0,0)}\oplus  \frak m_{(1,-1)}$ is a
trivial $M$-module. Consequently, $\frak m$ is a $(\frak g
\frak l_2, M)$-module, i.e., $\frak m$ is both a $\frak g
\frak l_2$- and an $M$-module and these actions commute.

The Lie algebra $\frak m$ is a generalized Kac-Moody algebra (see
\cite{B3}, where the definition used there allows that $\frak m$ might be
a generalized Kac-Moody algebra with some possibly nontrivial ``outer''
derivations adjoined, and see also \cite{Jur2}).
In fact, $\frak m$ is shown in \cite{Jur2} to be isomorphic to $\frak g(A)/
\frak c$, where
$\frak g(A)$ is the generalized Kac-Moody algebra associated to the
following matrix $A$ and $\frak c$ is the full center of $\frak g(A)$:

$${A} = \left( \begin{array}{c|c|c|c}
          2  & \begin{array}{ccc}
          0&\cdots&0    \end{array}
            & \begin{array}{ccc}   -1 & \cdots & -1  \end{array} & \cdots \\
         \hline
         \begin{array}{c}  0 \\ \vdots \\ \ 0  \end{array}
                      & \begin{array}{ccc} -2 & \cdots &-2 \\
                                         \vdots & \ddots & \vdots \\
                                       -2 & \cdots & -2   \end{array} &
                      \begin{array}{ccc} -3 & \cdots &-3 \\
                                         \vdots & \ddots & \vdots \\
                                       -3 & \cdots & -3
                                        \end{array} & \cdots\\
   \hline
          \begin{array}{c}  -1 \\ \vdots \\ -1  \end{array} &
          \begin{array}{ccc} -3 & \cdots &-3 \\
                                         \vdots & \ddots & \vdots \\
                                       -3 & \cdots & -3
                                        \end{array} &
          \begin{array}{ccc} -4 & \cdots &-4 \\
                                         \vdots & \ddots & \vdots \\
                                       -4 & \cdots & -4
                                        \end{array} & \cdots \\
          \hline
          \vdots & \vdots & \vdots
         \end{array}   \right)  .   $$

Consider the index set ${\cal I}_0 = \{ -1, 1, 2, 3, \cdots\}$. We shall
take our index set $I$, indexing the generators and simple roots of
$\frak g(A)$, to be the set $\cal I$ of pairs $(i,k) \in {\cal I}_0 \oplus
\Bbb Z_+$ with $1 \leq k \leq c(i)$, reflecting
the block form of the matrix. Thus the block
in position $(i,j) \in {\cal I}_0 \times {\cal I}_0$ has size $c(i)
\times c(j)$, and
entries $-(i +j)$ for $i, j \in {\cal I}_0$.
Then $\frak g(A)$ has generators
$ e_{ik} , h_{ik} , f_{ik}$ and simple roots
$\alpha_{ik}$, $(i,k) \in {\cal I}$. We will write
 $ e_{-1,1}$ as $e_{-1}$, $f_{-1,1}$ as $f_{-1}$ and $h_{-1,1}$
as $h_{-1}$.
Note that the simple roots
$\alpha_{ik}$ for fixed $i\in {\cal I}_0$,
$1\leq k \leq c(i)$, all have the same values on the Cartan subalgebra
$\frak h
\subset \frak h^e$.
Under the isomorphism
between $\frak g(A)/\frak c$ and $\frak m$, the subalgebras $\frak n^{\pm}
\subset
\frak g(A)$ are identified
with the subalgebras $\frak n^{\pm} \subset \frak m$ and
the elements $h_{ik}$ for
$(i,k) \in {\cal I}$ map to
$\l( {1-i \over 2}\r) h + \l( {1+i \over 2}\r)z \in \frak m_{(0,0)}$.
The $\Bbb Z \oplus \Bbb Z$-grading of the Lie
algebra $\frak m$ given by (\ref{eq:ng}) is determined by the root
space grading of $\frak m$, where the elements of
$\frak m^{\alpha_{ik}}$ are given degree $(1,i)$ for $1
\leq k \leq c(i)$.

The Lie algebra $\frak g(A)$ satisfies the hypothesis of
Theorem~\ref{thm:free}. Therefore, setting
\begin{equation}
{\scr V}_i= \coprod_{1 \leq k\leq c(i)} {U}( \Bbb C f_{-1})
\cdot f_{ik}, \label{eq:F}
\end{equation}
\begin{equation}
{\scr V}'_i= \coprod_{1 \leq k\leq c(i)} {U}( \Bbb C e_{-1})
\cdot e_{ik} \label{eq:E}
\end{equation}
for $i >0$ and defining
\begin{equation}
{\scr V} = \coprod_{i >0}{\scr V}_i \mbox{ and }
{\scr V}' = \coprod_{i >0}{\scr V}'_i,\label{eq:ik}
\end{equation}
it follows (see \cite{Jur2}) that
\begin{equation}
\frak m = \frak u^+ \oplus \frak g \frak l_2 \oplus \frak u^- ,
\label{eq:free} \end{equation}
with $ \frak u^- = L({\scr V})$ and $\frak u^+ = L({\scr V}')$.
In terms of the decomposition (\ref{eq:ng}),
$$
\frak u^+ = \coprod_{m,n >0} \frak m_{(m,n)},
$$
\begin{equation}
\frak u^- = \coprod_{m,n <0} \frak m_{(m,n)}. \label{eq:um}
\end{equation}
Note that
the subspaces $\frak u^{\pm}$ are $( \frak g \frak l_2, M)$-modules.

\begin{lemma}
Let ${\scr V}$, ${\scr V}_i$ and $\frak u^\pm$ be as above.
The $\Bbb Z \oplus \Bbb Z$-graded vector spaces ${\scr V}$, ${\scr
V}_i$ and $\frak u^\pm$ are
$(\frak g \frak l_2, M)$-modules.
For $i >0$, let $W_i$ denote the (unique up to isomorphism)
irreducible $\frak g
\frak l_2$-module of dimension $i$ on which $z$ acts as $i +1$.
Then ${\scr V}_i \cong W_i \otimes V^\natural_i$, where $W_i$ is regarded
as a trivial $M$-module and $V^\natural_i$ as a trivial $\frak g
\frak l_2$-module. Furthermore, as $M$-modules, ${\scr V}_i \cong
i{V}^\natural_i$ and, for $m,n >0$,
\begin{equation}
({\scr V}_i)_{(-m,-n)}\cong \l\{ \begin{array}{ll}
                 V^\natural_{i} & \mbox{ if } m+n =i+1\\
                  0 & \mbox{ otherwise.} \end{array}\r.
\end{equation}
\label{lem:V}
\end{lemma}

\noindent{Proof}:
Clearly, the space of highest weight vectors of ${\scr V}_i$
has basis $\{f_{ik} \ | \ 1 \leq k \leq c(i)\}$ and each $f_{ik}$ has
degree $(-1,-i)$.
Since $h$ acts on each $f_{ik}$ as multiplication by $i -1$ and $z$ acts as
multiplication by $i +1$,
each of these highest weight vectors generates a submodule isomorphic
to $W_i$. Therefore, $${\scr V}_i \cong W_i \otimes \coprod_{1 \leq k\leq
c(i)} \Bbb C f_{ik}.$$
The isomorphism in \cite{Jur2} between $\frak g (A)
/\frak c$ and $\frak m$ is such that the subspace $ \coprod_{1 \leq k
\leq c(i)} (\frak g (A)/\frak c)^{\alpha_{ik}}$
is mapped isomorphically to
$\frak m_{(1,i)}$, so the images of the $f_{ik}$ (respectively, the
$e_{ik}$) for
$1 \leq k\leq c(i)$ form a basis for the space
$\frak m_{(-1, -i)}$ (respectively, $\frak m_{(1,i)}$).
Identifying the space $\frak m_{(-1, -i)}$ with the $M$-module
$V^\natural_i$ via Lemma \ref{lem:ng}, we have
\begin{equation}
\coprod_{ 1 \leq k\leq c(i)} \Bbb C f_{ik} \cong V^\natural_i. \label{eq:fi}
\end{equation} In this way we give the ${\scr V}_i$ (and therefore ${\scr V}$)
$M$-module structures, so they are $(\frak g \frak l_2, M)$-modules.
$\square$ \vspace{1ex}

Just as in the case of the copy of $\frak g \frak l_2 =
\mbox{span}\,\{e,f,h,z\}$ inside $\frak m= P_1 / R$ described above,
it is easy
to identify copies of the vector spaces $\scr V$ and $\scr V'$ inside
$\frak m$: For $i >0$, $\frak m_{(-1,-i)}$ consists of
highest weight vectors for $\frak g\frak l_2$. The sum of all of the
$\frak g \frak l_2$-modules generated by the $\frak m_{(-1,-i)}$, $i
>0$, is a copy of $\scr V$ contained in $\frak m$. The space $\scr V'$ can
be analogously identified in $\frak m$. We know that the Lie
subalgebras of $\frak m$ generated by $\scr V$ and $\scr V'$ are free
Lie algebras. This suggests the following:

 \vspace{1ex}
\noindent{\bf Conjecture}: The associative algebra generated by the
set $\{v_0 \ |\ v \in \scr V\}$ of weight-zero components of the
vertex operators $Y(v,z)$ ($v \in \scr V$) and the associative algebra
generated by $\{v_0 \ | \ v \in \scr V'\}$, acting on $\frak m$,
are free associative algebras. Equivalently, these algebras are
isomorphic to $T(\scr V)$ and $T(\scr V')$, respectively.
 \vspace{1ex}

The Lie algebra $\frak m$ is not the only quotient of
$\frak g(A)$ which has an action of $M$. We now construct another
example. Let $\frak h_i$, $i \in \Bbb
Z_+$, be the span of the $h_{ik}$ for $1\leq k \leq
c(i)$. Let $\frak c_i = \frak c \cap \frak h_i$. Note that if $1 \leq
k,l\leq c(i)$, then $h_{ik} -h_{il} \in \frak c_i$ (since the
corresponding columns of $A$ are identical), and in fact these
elements span $\frak c_i$, so that $\dim \frak h_i/ \frak c_i =1$.
Define $\frak c' =
\coprod_{i =1}^\infty \frak c_i$, which is a central ideal of
$\frak g(A)$.

\begin{proposition}
The action of $M$ on the Lie algebra $ \frak m$ lifts naturally to an
action of $M$ on the Lie algebra $\frak g(A) / \frak c'$, in such a
way that $M$ acts trivially on $\frak h / \frak c'$.
 \label{prop:c}
\end{proposition}

\noindent Proof: We identify the elements $e_{ik}$ and $f_{ik}$,
$(i,k) \in {\cal I}$, in the
Lie algebras $\frak g(A)$, $\frak g(A)/ \frak c'$ and $\frak m$. The
invariant bilinear form on $\frak g(A)$ is preserved, in the obvious
sense, when we take
quotients of $\frak g(A)$ by central ideals. By
(\ref{eq:fi}), for each $i > 0$ we have that the span of the $f_{ik}$
and the span of
the $e_{ik}$, $ 1\leq k \leq c(i)$, are $M$-modules.
Recall that the $e_{-1}$ and $f_{-1}$ are fixed by the action of $M$.
Since $h_{ik} -h_{il} \in \frak c_i$ for $1 \leq k,l\leq c(i)$,
we can write the image of each $h_{ik}$ in $\frak g(A)/\frak c'$ as $h_i$.
The Lie algebra $\frak g(A)/ \frak c'$ then has a presentation with
generators $e_{ik},
f_{ik} , h_i$, $i \in {\cal I}_0$, $1 \leq k \leq c(i)$ and relations
induced by the relations (R1)-(R5) (the relation (R6) is vacuous) of
$\frak g(A)$:

\begin{description}
 \item[(R1)] $\left[ h_{i}, h_{j}\right] =0$,
 \item[(R2)] $\left[ h_{l}, e_{ik}\right]= -(l+i) e_{ik}$,
 \item[(R3)] $\l[h_{l}, f_{ik}\r]  =  (l+i) f_{ik}$,
 \item[(R4)] $\l[e_{ik} , f_{jm} \r] = (e_{ik}, f_{jm}) h_{i} $,
 \item[(R5)] $ (\ad e_{-1})^{i}e_{ik} =0$ and $(\ad f_{-1})^{i}f_{ik}=0 $
\end{description}
for $i, j \in {\cal I}_0$, $1 \leq k \leq c(i)$,  $1 \leq m \leq
c(j)$.

In $\frak g(A)/\frak c'$ let $M$ act trivially on $\frak h/
\frak c'$, so that $M$ acts trivially on all the $h_i$. Relations
(R1), (R2), (R3) and (R5) are $M$-invariant by inspection. The
relation (R4) is equivalent to the assertion that for all $e \in
\coprod_{ 1 \leq k\leq c(i)} \Bbb C e_{ik}$
and $f \in \coprod_{ 1 \leq k\leq c(i)} \Bbb C f_{ik}$,
$$[e,f] = (e,f)h_i.$$
The form $(\cdot, \cdot)$ is $M$-invariant on
$\coprod_{ 1 \leq k\leq c(i)} \Bbb C e_{ik} \times \coprod_{ 1 \leq
k\leq c(i)} \Bbb C f_{ik} $, because by Lemma \ref{lem:ng}, the form
is $M$-invariant on $\frak m$. Since $(e_{ik}, f_{lm}) = 0$ if $i \neq
l$, we conclude that relation (R4) is also $M$-invariant.
$\square$ \vspace{1ex}

Of course, $M$ also acts naturally on those generalized Kac-Moody
algebras formed by taking further quotients of $\frak g(A) / \frak c'$
by central ideals.

\subsection{Local Lie algebras}

We now review some facts about local Lie algebras. These will be used
(in Section 3) in the construction of certain generalized Kac-Moody
algebras. The material in this subsection is applicable for an
arbitrary field of characteristic zero.

\vspace{1ex}
\noindent {\bf Definition}: A {\em local Lie algebra} is a graded vector
space $\frak a = \frak a_{-1} \oplus \frak a_0 \oplus \frak a_1$ with
skew-symmetric bilinear
maps $[\cdot ,\cdot ]_{loc}:\frak a_i \times \frak a_j \rightarrow
\frak a_{i+j}$
for $|i|,|j|, |i+j| \leq 1$, satisfying:
\begin{enumerate}
\item[i.] $[\cdot ,\cdot ]_{loc}:\frak a_0 \times \frak a_0
\rightarrow \frak a_{0}$ gives $ \frak a_0$ the structure of a Lie algebra.
\item[ii.] $[\cdot ,\cdot ]_{loc}:\frak a_0 \times \frak
a_{\pm 1}\rightarrow \frak a_{\pm 1}$ gives $\frak a_{\pm 1}$ the structure
of an $\frak a_0$-module.
\item[iii.] $[\cdot ,\cdot ]_{loc}:\frak a_{-1} \times \frak a_1
\rightarrow \frak a_0$ gives a map $\frak a_{-1} \otimes \frak a_{1}
\rightarrow \frak a_0$ of $\frak a_0$-modules.
\end{enumerate}

Let $\frak l = \coprod_{i \in \Bbb Z} \frak l_i$ be a $\Bbb Z$-graded
Lie algebra
(with product $[\cdot ,\cdot ]$). Define $[\cdot ,\cdot ]_{loc}:
{\frak l_i} \times
{\frak l_j} \rightarrow {\frak l_{i+j}}$ to be the
restriction of $[\cdot ,\cdot ]$ to $\frak l_i \times \frak l_j$ whenever
$|i|,|j|,|i+j| \le 1$. Then $\frak l_{-1} \oplus \frak l_0 \oplus \frak
l_1$ has the structure of a local Lie algebra, which we call the {\em
local part of} $ \frak l$ and denote by
$\frak l_{loc}$.

We say that two local Lie algebras ${\frak a}, {\frak b}$ are {\em
isomorphic}
 if there is an isomorphism $\phi: {\frak a} \rightarrow {\frak
b}$ of
graded vector spaces such that $\phi [a_i, b_j]_{loc} = [\phi a_i,\phi
b_j]_{loc}$ for all $a_i \in {\frak a_i}, b_j \in {\frak a_j},
|i|,|j|,|i+j| \le 1$. A homomorphism of local Lie algebras is defined
analogously.

\vspace{1ex}
\noindent {\bf Definition}: A $\Bbb Z$-graded Lie algebra $\frak l =
\coprod_{i \in {\Bbb Z}}\frak l_i$ is {\em associated to the local Lie
algebra $\frak a=
\frak a_{-1} \oplus \frak a_0 \oplus \frak a_1$} if $\frak l_{loc}$ is
isomorphic to $\frak a$,
 and $\frak l$ is generated by $\frak l_{-1} \oplus
\frak l_0 \oplus \frak l_1$.
\vspace{1ex}

The following result is due to Kac \cite[ Propositions 4 and 7]{K1}
in case $\frak a$ is finite-dimensional over a field of characteristic
zero. The same proof remains valid for arbitrary $\frak a$.

\begin{proposition} Let $\frak a = \frak a_{-1} \oplus \frak a_0 \oplus
\frak a_1$ be a local Lie algebra. Then up to isomorphism there is a
unique $\Bbb Z$-graded Lie algebra
$\frak a_{max} = \coprod_{i \in \Bbb Z} \frak a_i$ which is associated
to $\frak a$ and is maximal in the sense that any graded Lie algebra associated
to $\frak a$ is a quotient of $\frak a_{max}$.  Furthermore,
$\frak a_{max}^{\pm}$ is the free Lie algebra $L(\frak a_{\pm 1})$
(recall the notation in Section 2.1).
If $(\cdot ,\cdot )_{loc}$ is a symmetric bilinear form on $\frak a$
which is
invariant with respect to $[\cdot ,\cdot ]_{loc}$, in the obvious
sense, then $(\cdot ,\cdot)_{loc}$ has a
unique extension to an invariant symmetric bilinear form $(\cdot ,\cdot )$ on
$\frak a_{max}$.
\label{thm:kac}
\end{proposition}

\section{Construction of a Lie algebra from a Kac-Moody
algebra and a module}

In this section we will give a construction of a Lie algebra starting
{}from a Kac-Moody algebra and a suitable module for this algebra.  We will
show that every generalized Kac-Moody algebra in which no pair of
distinct imaginary simple roots is orthogonal arises from our
construction (so that in particular, the Monster Lie algebra does).

Let  $\Gamma$ be an index set, and let $V = \coprod_{\gamma \in \Gamma}
V_{\gamma}$ be a $\Gamma$-graded vector space over
$\Bbb F$, which for the general considerations in this section,
through Corollary 3.1, we take to be an arbitrary field.  For a subset
$S \subset \Gamma$
let $$V_{(S)} = \coprod_{\gamma \in \Gamma \backslash S} V_{\gamma}.$$ Clearly,
$V_{(S)} \cap V_{(T)} = V_{(S \cup T)}$ and so $\{V_{(S)} | S \subset
\Gamma, S {\rm \ finite}\}$ forms a base of neighborhoods of $0$ for a
topology on $V$.  As usual we give the base field the discrete
topology.  Let $V' \subset V^*$ denote the space of continuous linear
functionals on $V$.  Note that the topology of $V$ and hence the space
$V'$ depend on the grading of $V$.
Using a given involution of $\Gamma$, i.e., a map
$$\Gamma \longrightarrow \Gamma$$
$$\gamma \longmapsto \gamma'$$
of order one or two, we define a $\Gamma$-grading on $V'$ by
identifying $(V_\gamma)^*$ with
$$V'_{\gamma'} = \{v' \in V^* | v'(v) = 0 {\rm \ for \ }
v \in V_{(\{\gamma\})} \}$$
for $\gamma \in \Gamma$; then
$$V' = \coprod_{\gamma \in \Gamma} V_{\gamma}'.$$
(Later, we shall take $\Gamma$ to be an abelian group and $\gamma'$ to
be $-\gamma$.)

We say that
a bilinear form $(\cdot ,\cdot)$ on $V$ is {\em compatible} with the grading
and with a given involution $\gamma \mapsto \gamma'$ of $\Gamma$
if $(V_{\gamma},V_{\mu}) = 0$ if $\mu \ne
\gamma'$. We say that $V$ is {\em finite-dimensionally graded} if each
$V_{\gamma}$ is finite-dimensional.

Suppose that
$(\cdot ,
\cdot)$ is a nondegenerate (symmetric) bilinear form on $V$ which is
compatible with the grading and with a given involution $\gamma
\mapsto \gamma'$.
Then the restriction of $(\cdot , \cdot)$ to each $V_{\gamma}
\times V_{\gamma '}$ is nondegenerate.   It follows that if $V$ is
finite-dimensionally graded, then for $f \in V'$
there exists a unique $v_f \in V$ such that
$$f(v) = (v_f,v)$$
for all $v \in V$.

\begin{lemma} Let ${\frak a}_0$ be a Lie
algebra finite-dimensionally graded by $\Gamma$ (as a vector space) and
let ${\frak a} =  {\frak a}_{-1}
\oplus {\frak a}_0 \oplus {\frak a}_1$ be an ${\frak a}_0$-module
graded by $\{-1, 0,1\}$.
Let $(\cdot ,\cdot )$ be an ${\frak a}_0$-invariant symmetric bilinear
form on ${\frak a}$.  Assume that $({\frak a}_i,{\frak a}_j) = 0$
whenever $i+j \ne 0$, that the restriction of $(\cdot,\cdot)$ to
$\frak a_0 \times \frak a_0$ is nondegenerate
and compatible with the grading on $\frak a_0$ and with a given
involution on $\Gamma$, and that
for any $a_{\pm 1} \in {\frak a_{\pm 1}}$ the
linear functional $$a_0 \longmapsto (a_{-1},a_0 \cdot a_1)$$ is
continuous on $\frak a_0$.   Then there is a unique local Lie algebra
structure on
${\frak a}$ which extends the ${\frak a}_0$-module structure of
${\frak a}$ and such that $(\cdot ,\cdot )$ is invariant with respect
to this structure. \label{lem:loc}

\end{lemma}

\noindent Proof:  Let $a_i \in {\frak a}_i$ for $i = -1,0,1$.
We (must) define
$$[a_0,a_{\pm 1}]_{loc} = - [a_{\pm 1},a_0]_{loc} = a_0\cdot a_{\pm
1}.$$
The rest of $[\cdot,\cdot]_{loc}$ is determined uniquely as
follows: By
hypothesis, the linear functional
$$\begin{array}{l}
{\frak a}_0 \longrightarrow \Bbb F\\
a_0 \longmapsto (a_{-1},[a_1,a_0]_{loc})
\end{array}$$
is continuous.  Since $(\cdot,\cdot)$ is nondegenerate and compatible
with the grading on ${\frak a_0}$
there is a unique element of $\frak a_0$, which we
denote by $[a_{-1},a_1]_{loc}$, such that
\begin{equation}
([a_{-1},a_1]_{loc},a_0) = (a_{-1},[a_1,a_0]_{loc})\label{eq:inv}
\end{equation}
for all $a_0 \in {\frak a}_0$.  Clearly, $[\cdot,\cdot]_{loc}$ is
bilinear on ${\frak a_{-1}} \times {\frak a_1}$. Set $[a_1,a_{-1}]_{loc} = -
[a_{-1},a_1]_{loc}$.

Then for any $a_0,a'_0 \in {\frak a}_0$ we have
\begin{eqnarray*}
([[a_{-1},a_{1}]_{loc},a_0]_{loc},a'_0)&=& ([a_{-1},a_{1}]_{loc},
[a_0,a'_0]_{loc}) =
                  (a_{-1},[a_{1},[a_0,a'_0]_{loc}]_{loc})\\
  & =& (a_{-1},[[a_{1},a_0]_{loc},a'_0]_{loc}) +
         (a_{-1},[a_0,[a_{1},a'_0]_{loc}]_{loc}) \\
  & =& ([a_{-1},[a_{1},a_0]_{loc}]_{loc},a'_0) +
             ([a_{-1},a_0]_{loc},[a_{1},a'_0]_{loc}) \\
  & =& ([a_{-1},[a_{1},a_0]_{loc}]_{loc},a'_0) +
           ([[a_{-1},a_0]_{loc},a_{1}]_{loc},a'_0).
\end{eqnarray*}
In view of the
nondegeneracy of $(\cdot ,\cdot )|_{{\frak a}_0  \times   {\frak
a}_0}$ this implies
$$[[a_{-1},a_{1}]_{loc},a_0]_{loc} = [a_{-1},[a_{1},a_0]_{loc}]_{loc}
     + [[a_{-1},a_0]_{loc},a_{1}]_{loc}.$$
This completes the proof that $[\cdot ,\cdot ]_{loc}$ gives ${\frak
a}$ the structure of
a local Lie algebra.  The invariance of $(\cdot ,\cdot )$ now follows
{}from (\ref{eq:inv}) together with the ${\frak a}_0$-invariance of
$(\cdot ,\cdot)$, the skew symmetry of $[\cdot,\cdot]_{loc}$ and the
symmetry of $(\cdot,\cdot)$.
$\square$ \vspace{1ex}

Now let $W = \coprod W_i$ be
a direct sum of finite-dimensionally graded
vector spaces.  For each $i$, identify $W_i^*$ with $\{w^* \in W^* |
w^*(w) = 0 {\rm \ for
\ } w\in W_j, j \ne i\}$.  Then $W'_i \subset W_i^*$ (defined above)
is identified with a subspace of $W^*$.  We set
$$W' = \coprod W'_i \subset W^*.$$
For $w \in W, w' \in W'$ define $(w',w) = (w,w') = w'(w)$.  Also set
$(W,W) = (W',W') = 0$.  Thus $(\cdot ,\cdot )$ is a nondegenerate symmetric
bilinear form on $W \oplus W'$.

\begin{corollary}  Let $\frak a_0$ be a Lie algebra which is
finite-dimensionally graded, as a Lie algebra, by an abelian group
$\Gamma$. Assume that $\frak a_0$ has a nondegenerate
invariant symmetric bilinear form $(\cdot ,\cdot )$
compatible with the grading and the involution $\gamma \mapsto -\gamma$
of the abelian group $\Gamma$. Let $W = \coprod W_i$ be a
direct sum of finite-dimensionally $\Gamma$-graded
$\frak a_0$-modules. Then there is a unique $\frak a_0$-module
structure on $W'$ such that the nondegenerate symmetric bilinear form
$(\cdot, \cdot)$ on $W \oplus W'$ defined above is $\frak
a_0$-invariant, and $W'$ is $\Gamma$-graded as an $\frak a_0$-module.
Set $\frak a_1 = W, \frak a_{-1} = W', \frak a = \frak a_{-1} \oplus
\frak a_0 \oplus \frak a_1$.  Let $(\cdot ,\cdot )$ be the nondegenerate
$\frak a_0$-invariant symmetric bilinear form on $\frak a$ which
extends the forms $(\cdot ,\cdot )$ on $\frak a_0$ and on $W \oplus
W'$,
so that $(\frak a_i,\frak a_j) = 0$
if $i+j \ne 0$, and $(a_{-1},a_1) = a_{-1}(a_1)$ for
$a_{\pm 1} \in \frak a_{\pm 1}$.  Then $\frak a$ has a unique local
Lie algebra structure extending the $\frak a_0$-module structure of
$\frak a$ such that $(\cdot ,\cdot )$ is invariant. \label{cor:loc}
\end{corollary}

\noindent Proof: To prove the first assertion, we first note that for
each $i$, the bilinear form $(\cdot ,\cdot )$ on $W_i \oplus W_i'$ is
compatible with the grading and the involution $\gamma \mapsto
-\gamma$. For $a_0 \in \frak a_0$ and $w_i' \in W_i'$, the linear
functional
$$\varphi : w_i \mapsto (w_i',a_0\cdot w_i)$$
is continuous. In fact, it is sufficient to show this for $a_0$ in some
$(\frak a_0)_\mu$ and $w_i'$ in some $(W_i')_\nu$, in which case
$\varphi$ vanishes on $(W_i)_\gamma$ unless $\gamma + \mu + \nu =0$. Thus
$\varphi$ is continuous. We (must) define $a_0 \cdot w_i' \in W_i'$ so
that
$$(a_0 \cdot w_i', w_i) = -(w_i',a_0 \cdot w_i),$$
and $a_0 \cdot w_i'$ is clearly bilinear in $a_0$ and $w_i'$, and
clearly defines an $\frak a_0$-module action on $W'$. It is
straightforward to verify that $W'$ is $\Gamma$-graded as an
$\frak a_0$-module. This completes the proof of the first assertion.

To prove the remaining assertion, we use Lemma \ref{lem:loc}. To apply
this, we
must show that the linear functional
$$\psi: a_0 \longmapsto (a_{-1},a_0 \cdot a_1)$$
is continuous for all $a_{\pm 1} \in {\frak a}_{\pm 1}$.  It is
sufficient to verify this for $a_1$ in some $(W_i)_\mu$ and $a_{-1}$
in some $(W'_j)_\nu$.  Then $\psi$ vanishes on ${\frak a}_{\gamma}$
unless $\gamma + \mu + \nu = 0$ and hence $\psi$ is continuous, as
required. $\square$ \vspace{1ex}

Now we return to the setting of Section 2.2. Let $S$ be a finite set
and let $B = (b_{ij})_{i,j \in S}$ be a
matrix satisfying the
conditions (C1)-(C3) and the additional condition
$b_{ii} >0$ for all $i \in S$.
Then $\frak g(B)^e$ is the (symmetrizable) Kac-Moody algebra
associated to $B$.
Recall that the symmetric invariant bilinear form
$(\cdot ,\cdot)_{\frak g(B)^e}$ on $\frak g(B)^e$ is nondegenerate and
satisfies the condition $(e_i, f_j)_{\frak g(B)^e} =\delta_{ij}$ for
all $i,j \in S$.

If $V$ is an (irreducible) integrable
lowest weight $\frak g(B)^e$-module (necessarily finite-dimensionally
graded by its weight-space decomposition) with
lowest weight $\lambda$, then $V'$ carries the structure of an integrable
highest weight module with highest weight $-\lambda$.

Let $\frak z$ be a finite-dimensional abelian Lie algebra and let
$\frak a_0 = \frak g(B)^e \oplus \frak z$, a direct sum of Lie
algebras.

Let $W = \coprod_{j \in J} W_j$ be a direct sum of $\frak a_0$-modules,
where each $W_j$ is an integrable lowest weight
$\frak g(B)^e$-module, with lowest weight $\mu_j$ and lowest weight
vector $e_j$. (We assume that the two index sets $S$ and $J$ are
disjoint.)
Since each element of $\frak z$ must act as a scalar on $W_j$, there
is some $\gamma_j \in {\frak z}^*$ such that $z \cdot w =
\gamma_j(z)w$ for all $z \in \frak z,\ w \in W_j$. Assume for
convenience that
$$\bigcap_{j \in J}\mbox{Ker\,}\gamma_j =0.$$

Let
$(\cdot ,\cdot )_{\frak z}$ be a nonsingular symmetric bilinear form on
$\frak z$ (which is necessarily $\frak z$-invariant).
Define $(\cdot ,\cdot )$ on
$\frak a_0$ by $(g + z, g' + z') = (g,g')_{\frak g(B)^e} -
(z,z')_{\frak z}$ for $g,g' \in \frak g(B)^e, z,z' \in \frak z$. Note
that this form is symmetric and nondegenerate.

Now $W' = \coprod_{j \in J}W'_j$ (see above) and each $W'_j$ is an
irreducible integrable highest weight $\frak g(B)^e$-module, with highest
weight $- \mu_j$ and a highest weight vector $f_j$ which we choose so
as to satisfy the condition $(e_j,f_j) =
1$.  Clearly, $z \cdot w' = - \gamma_j (z) w'$ for all $z \in \frak z$
and $w' \in W_j'$. Also, $(e_i, f_j) =\delta_{ij}$ for $i,j \in J$.

Set $\frak k = \frak h^e \oplus \frak z$ and note that $(\cdot ,\cdot )$ is
nonsingular on $\frak k$.  We identify ${\frak k}^*$ with
$(\frak h^e)^* \oplus
{\frak z}^*$. Let $(\cdot ,\cdot )_{\frak h^e}$ denote the restriction
of the form $(\cdot ,\cdot )_{\frak g(B)^e}$ to $ \frak h^e$.
Then for $\mu \in (\frak h^e)^*$ there exists a unique $h_ \mu \in
\frak h^e$ such
that $(h_ \mu , h)_{\frak h^e} = \mu (h)$ for all $h \in \frak h^e$
and for $\gamma \in\frak z^*$
there exists a unique $z_ \gamma \in \frak z$ such that $(z_
\gamma, z)_{\frak z} = \gamma (z)$ for all $z \in \frak z$.
We denote $h_ \mu +
z_ \gamma$ by $k_{\mu - \gamma} \in \frak k$ and note that $(k_{\mu -
\gamma},k) = (\mu - \gamma)(k)$ for all $k \in \frak k$.  As usual, we may
transfer the nonsingular forms $(\cdot ,\cdot )_{\frak h^e}$ to
$(\frak h^e)^*$ and $(\cdot ,\cdot )_{\frak z}$ to $\frak z^*$
by defining
$(\alpha ,\beta)_{(\frak h^e)^*} = (h_\alpha, h_\beta)_{\frak h^e}$
for $\alpha ,\beta \in (\frak h^e)^*$ and  $(\gamma,\sigma)_{\frak z^*} =
(z_{\gamma}, z_{\sigma})_{\frak z}$ for $\gamma, \sigma \in  \frak z^*$.
There is a corresponding form $(\cdot ,\cdot)$ on
$\frak k^*$ defined by $(\cdot ,\cdot)_{(\frak h^e)^*}-
(\cdot,\cdot)_{\frak z^*}$. Thus
$(\nu,\tau) = (k_ \nu, k_ \tau)$ for $\nu, \tau \in  \frak k^*$. Note
that
$$\frak z = \mbox{span\,}\{ z_{\gamma_j} \ |\ j \in J\}.$$

With the $\frak a_0$ given above and $\Gamma = \frak k^*$, let
$${\frak a} = W' \oplus \frak a_0 \oplus W = \frak a_{-1} \oplus \frak a_0
\oplus \frak a_1$$
be the local Lie algebra given by Corollary \ref{cor:loc}
and let ${\frak a}_{max}$ be the associated $\Bbb Z$-graded Lie
algebra given by Proposition \ref{thm:kac}.

Next we give some basic bracket relations together with relevant
definitions:

\begin{proposition} The following relations hold in $\frak a$:
\begin{enumerate}

\item[(a)] If $i,j\in S$ then $[e_i,f_j] =\delta_{ij} h_i =
\delta_{ij} h_{\alpha_i}$.

\item[(b)] If $i,j \in J$ then $[e_i,f_j] = \delta_{ij}k_i$, where
$$k_i =k_{\mu_i +\gamma_i}.$$

\item[(c)] If $i \in S$ and $j \in J$ then $[e_i,f_j] = 0$ and
$[f_i,e_j] =0$.

\end{enumerate}
\end{proposition}

\noindent Proof:  (a) is clear.

(b) For $i,j \in J$, $[e_i,f_j]$ is an element of $\frak a_0$ which is
orthogonal to the subalgebras $\frak n^\pm \subset \frak g(B)^e$, so
that $[e_i,f_j] \in \frak k$. Now
let $k \in \frak k$.  Then $(k,[e_i,f_j]) =
([k,e_i],f_j) = ((\mu_i(k) + \gamma_i(k))e_i,f_j) =
\delta_{ij}(\mu_i(k) + \gamma_i(k))$, so $[e_i,f_j] =
\delta_{ij}k_{\mu_j + \gamma_j}$.

(c) follows from the definitions of $e_j$ and $f_j$.
$\square$ 

\begin{proposition} The following relations hold in $\frak a$:
\begin{enumerate}

\item [(a)] Let $i,j \in S$.  Then $[h_i,e_j] =
a_{ij}e_j$ and $[h_i,f_j] = -a_{ij}f_j$, where $$a_{ij} = b_{ij} =
(\alpha_i,\alpha_j).$$

\item [(b)]  Let $i \in S, j \in J$.  Then
$\l[h_i,e_j\r] = a_{ij}e_j$,
$\l[h_i,f_j\r] = -a_{ij}f_j$,
$\l[k_j,e_i\r] = a_{ji}e_i$, and
$\l[k_j,f_i\r] =  -a_{ji}f_j ,$
where $$a_{ij} = a_{ji}=(\alpha_i,\mu_j).$$

\item [(c)] Let $i,j \in J$.  Then $[k_i,e_j] = a_{ij}e_j$ and
$[k_i,f_j] = -a_{ij}f_j$, where
$$a_{ij} =
 (\mu_i,\mu_j)_{(\frak h^e)^*} - (\gamma_i,\gamma_j)_{\frak z^*}.$$

\end{enumerate}
\end{proposition}

\noindent Proof:  (a) is clear.

(b)  We have
$$[h_i,e_j] = \mu_j(h_i)e_j =
\mu_j(h_{\alpha_i})e_j = (\alpha_i,\mu_j)e_j$$ and
$$[k_j,e_i] = [k_{\mu_j + \gamma_j},e_i] =
\alpha_i(h_{\mu_j})e_i = (\mu_j,\alpha_i)e_i.$$
The other cases are similar.

(c)  We have
\begin{eqnarray*}
[k_i,e_j] &=& [k_{\mu_i + \gamma_i},e_j] = [h_{\mu_i}
- z_{\gamma_i},e_j] \\
&=& (\mu_j(h_{\mu_i}) - \gamma_j(z_{\gamma_i}))e_j =
((\mu_i,\mu_j)_{(\frak h^e)^*} - (\gamma_i,\gamma_j)_{\frak z^*})e_j,
\end{eqnarray*}
and the remaining case is similar.
$\square$ \vspace{1ex}

Let $A= (a_{ij})_{i,j \in S \cup J}$,
where the $a_{ij}$ are defined in the above proposition.
Let $\frak d_B$ be the derivation space associated with $\frak g(B)$,
i.e., $\frak g(B)^e = \frak d_B \ltimes \frak g(B)$.
Note that the containment
$\frak d \ltimes \frak g(A) \subset \frak g(A)^e$ can be strict. We
assume that the restriction $(\cdot, \cdot)$ to $\frak g(B)^e$ of the
symmetric bilinear form $(\cdot, \cdot)$ on $\frak g(A)^e$ has
all the usual properties, in particular, that it is nondegenerate.
The radical $\mbox{Rad}(\cdot, \cdot)$ of the form $(\cdot, \cdot)$ on
$\frak d_B \ltimes \frak g(A)$ is central in $\frak g(A)^e$.

\begin{theorem}\

(a)  Assume that $a_{ij}=(\mu_i,\mu_j)_{(\frak h^e)^*} -
(\gamma_i,\gamma_j)_{\frak z^*} \le 0$ for all $i,j
\in J$.  Then $A$ satisfies {\em  (C1)-(C3)} and the Lie
algebra $\l(\frak d_B \ltimes
{\frak g}(A)\r)/\mbox{\em Rad}(\cdot, \cdot)$ is a Lie algebra
associated to the local Lie algebra $\frak a$. In particular, this Lie
algebra is a quotient of ${\frak a}_{max}$.

(b) Assume that $(\mu_i,\mu_j)_{(\frak h^e)^*} -
(\gamma_i,\gamma_j)_{\frak z^*}  \le 0$ for all $i,j
\in J$ and that $(\mu_i,\mu_j)_{(\frak h^e)^*} -
(\gamma_i,\gamma_j)_{\frak z^*}  < 0$ for all $i \ne j$.
Then $(\frak d_B \ltimes \frak g(A))/\mbox{\em Rad}(\cdot, \cdot) \cong
{\frak a}_{max}$.

\end{theorem}

\noindent Proof:
We first prove (a).
Since $W_j$ is an integrable lowest weight module
with lowest weight $\mu_j$, the properties (C1)-(C3) are clear.

 We shall use the symbols
$e^A_i, f^A_i$ and $h^A_i$, $i \in S \cup J$, to denote the generators
of $\frak g(A)$,
and $\frak h^A$ will denote the Cartan subalgebra of $\frak g(A)$.
Using the notation defined in (\ref{eq:dec}) above, we have $\frak r=
\frak g(B) +  \frak h^A$ and
$\frak g(A) = \frak u^+ \oplus \frak r
\oplus \frak u^-$. Thus $\frak d_B \ltimes\frak g(A) = \frak u^+ \oplus
\frak r^e \oplus \frak u^-$, where $\frak r^e = \frak d_B \ltimes
\frak r$. We give the Lie algebra $\frak d_B \ltimes\frak
g(A)$ the $\Bbb Z$-grading determined by:
$$\deg e_j = -\deg f_j =\l\{\begin{array}{ll}
                             0 & \mbox{ if } j \in S \\
                             1 & \mbox{ if } j \in J. \end{array}\r.
$$
Note that $\mbox{Rad}(\cdot, \cdot) \subset (\frak d_B \ltimes\frak
g(A))_0$, so that we can identify the elements  $e^A_i$ and $f^A_i$
with their images in $(\frak d_B \ltimes\frak g(A))/\mbox{Rad}(\cdot,
\cdot) $. Let $\frak l = (\frak d_B \ltimes\frak g(A))/\mbox{Rad}(\cdot,
\cdot) $.
As $\frak d_B \cap \mbox{Rad}(\cdot,\cdot) = 0$, we can also identify
$\frak d_B$ with its image in
$\frak l$. The $\Bbb Z$-grading defined on $\frak d_B \ltimes\frak g(A)$
induces a $\Bbb Z$-grading on $\frak l$.

Then $\frak l_1 = \coprod_{j \in J} U(\frak n_S^+)e^A_j$, $\frak l_{-1} =
\coprod_{j \in J} U(\frak n_S^-)f^A_j$ and $\frak l_0 =
\frak r^e/\mbox{Rad}(\cdot, \cdot)$. The subspace
$\frak l_{loc}=\frak l_{-1} \oplus \frak l_0 \oplus \frak l_{1}$ is the
local part of the Lie algebra $\frak l$ ($\frak l_{loc}$ generates
$\frak l$ since it contains $\frak d_B$ and the
generators $e^A_i$ and $f^A_i$, $i \in S\cup J$).

Thus, by Proposition \ref{thm:kac}, to prove that $\frak l$ is a quotient
of $\frak a_{max}$ it is
sufficient to prove $\frak l_{loc}\cong \frak a$. By the uniqueness
assertion in Lemma \ref{lem:loc}, it is
enough to give a linear isomorphism
$$\varphi: \frak l_{loc} \rightarrow \frak a$$
preserving the gradings by the set $\{-1,0,1\}$, such that for $x \in
\frak l_0$, $y \in \frak l_{loc}$,
\begin{equation}
[\varphi(x),\varphi(y)] =\varphi ([x,y]) \label{eq:loc1}
\end{equation}
and such that for all $x,y \in \frak l_{loc}$,
\begin{equation}
(\varphi(x), \varphi(y)) = (x,y). \label{eq:loc2}
\end{equation}

To do this we will define a surjective homomorphism
of graded vector spaces, which we still call $\varphi$,
$$\varphi: \frak l_{-1} \oplus \frak r^e \oplus
\frak l_{1} \rightarrow \frak a_{-1}
\oplus \frak a_0 \oplus \frak a_{1},$$
and show that (\ref{eq:loc1}) and (\ref{eq:loc2}) are satisfied and that
the kernel of $\varphi$ is $\mbox{Rad}(\cdot, \cdot)$.

We define $\varphi$ as a linear map by prescribing
$\varphi$ on the subspaces $\frak l_{\pm 1}$, $\frak g(B)^e$ and the span
of the $h^A_j$ for $j \in J$.

Each $W_j$ for $j \in J$ is an integrable lowest weight
$\frak g(B)^e$-module of weight $\mu_j$ and so is isomorphic to the
integrable lowest weight $\frak g(B)^e$-module $U(\frak n_S^+)e^A_j$
of weight $\mu_j$ by an isomorphism $\varphi$ which takes the lowest weight
vector $e^A_j$ to $e_j$. This gives us
$\varphi:\frak l_{1}{\buildrel \sim \over \rightarrow} \frak a_1 $.
The isomorphism $\varphi:
\frak l_{-1} {\buildrel \sim \over \rightarrow} \frak a_{-1}$ is
defined analogously.

We define $\varphi$ to be the identity on $\frak g(B)^e$.

To define $\varphi$ on all of $\frak r^e$, all that remains
is to define $\varphi$ on $h^A_j \in \frak h^A$
with $j \in J$. Set $\varphi(h^A_j)= k_j=k_{\mu_j+\gamma_j}\in \frak k$.

This defines a homomorphism of graded vector
spaces $\varphi: \frak l_{-1}\oplus \frak r^e \oplus \frak l_1
\rightarrow \frak a$.
That $\varphi$ satisfies (\ref{eq:loc1})
follows from Proposition 3.4. This map is surjective because the
$z_{\gamma_j}$ span $\frak z$.

It is clear from the definition of the
forms that $(x,y) = (\varphi(x),\varphi(y))$ for $x,y \in
\frak g(B)^e$ . The invariant bilinear
forms on $\frak l_{-1}\oplus \frak r^e \oplus \frak l_1$ and on $\frak a$
satisfy the relations
$$(e^A_i,f^A_j)=(e_i,f_j)= \delta_{ij}$$ for all $i,j \in J$.
Since an invariant pairing between a lowest and highest weight $\frak
g(B)^e$-module is determined by its value on the pair consisting
of the lowest and highest weight vectors, we conclude that
$(x,y) = (\varphi(x),\varphi(y))$ for $x \in \frak l_{ 1}$ and
$y \in  \frak l_{- 1}$.
Furthermore, for $i \in S\cup J ,j \in J$
\begin{eqnarray*}
(h^A_i,h^A_j) = a_{ij}& =&\l\{ \begin{array}{ll}
           (\alpha_i,\mu_j+\gamma_j) & \mbox{ if } i\in S \\
          (\mu_i + \gamma_i,\mu_j+\gamma_j)& \mbox{ if } i\in J
               \end{array}\r. \\
                  & =& \l\{ \begin{array}{ll}
             (h_{\alpha_i}, k_{j})& \mbox{ if } i\in S \\
           (k_{i}, k_{j})& \mbox{ if }
               i\in J  \end{array}\r. \\
                  &= & (\varphi(h_i),\varphi(h_j)) .
\end{eqnarray*}
Also, it is clear that for $d \in\frak d_B$ and $j \in J$, $(d,h^A_j)
= (\varphi(d), \varphi(h^A_j))$.
It follows that for all $x,y \in \frak l_{-1}\oplus \frak r^e \oplus
\frak l_1$,
we have $(x,y) = (\varphi(x),\varphi(y))$.

The kernel of $\varphi$ clearly lies in $\frak d_B \oplus \frak h^A$.
Let $h$ be an element of the kernel.
Then for all $x\in \frak l_{-1}\oplus \frak r^e \oplus \frak l_1$,
$(h, x)=(\varphi (h), \varphi (x)) =0 $, so that $h \in
\mbox{Rad}(\cdot, \cdot)$. Conversely, if $y \in \mbox{Rad}(\cdot,
\cdot)$, then $(\varphi(y),a)=0$ for all $ a\in \frak a$,
since $\varphi$ is surjective.
As the bilinear form is nondegenerate on $\frak a_0$, this means that
$\varphi(y)=0$. We conclude that the kernel of the map $\varphi$ is
$\mbox{Rad}(\cdot, \cdot)$, completing the proof of (a).

Part (b) then follows by Theorem~\ref{thm:free} and the
definition of $\frak d_B \ltimes  \frak g(A)$. $\square$

\begin{remark}{\em
After possible reordering of rows and columns, every
finite rank matrix
satisfying  (C1)-(C3) and having only finitely many positive
diagonal entries arises as the matrix $A$ for a
suitable choice of the generalized Cartan matrix $B$, the module $W$
and the elements $\gamma_i$.
In fact, let $P =
(p_{kl})_{k,l \in K}$ be any matrix as indicated.
Let $S = \{i \in K | p_{ii} > 0 \}$ and $J = \{j \in K | p_{jj} \le 0
\}$, so that $S$ is a finite set. Reorder the rows and columns of $P$
so that the rows and columns
indexed by elements of $S$ precede those indexed by elements of $J$.
Let $B = (p_{kl})_{k,l \in S}$.  Consider the Kac-Moody algebra
$\frak g(B)^e$. For $j \in J$, define $\mu_j$ to be any element of
$ (\frak h^e)^*$ such that
$\mu_j (h_i) = p_{ij}$ for $i \in S$.  Let $W_j$ be the
integrable lowest weight $\frak g(B)^e$-module with lowest weight
$\mu_j$.  Define
$q_{ij} = (\mu_i,\mu_j) - p_{ij}$ for $i,j \in J$ and let $Q =
(q_{ij})_{i,j \in J}$.  Let ${\frak z}$ be an abelian Lie algebra of
(finite) dimension equal to the rank of $Q$, equipped with a nonsingular
symmetric bilinear form $(\cdot,\cdot)_{\frak z}$. Let
$(\cdot,\cdot)_{\frak z^*}$ be the
corresponding form on $\frak z^*$. Let $\frak t$ be a vector space over
$\Bbb C$ with basis $\{ \eta_j\ | \ j \in J\}$. Define an inner
product on $\frak t$ by $(\eta_i, \eta_j)= q_{ij}$ for all $i,j \in
J$. Then $\frak t/\mbox{Rad}(\cdot,\cdot)$ has dimension equal to rank
$Q$, hence to $\dim \frak z^*$. We may therefore identify
$\frak t/\mbox{Rad}(\cdot,\cdot)$ with $\frak z^*$. Let $\gamma_j$
denote the image of $\eta_j$ under this identification, so that
$(\gamma_i, \gamma_j)_{\frak z^*} = q_{ij}$ for all $i,j \in J$.
For these choices of $B,\ W = \coprod_{j \in J} W_j$ and the
$\gamma_j$, the resulting matrix $A$ is equal to $P$.} \end{remark}

We give such a construction of the Monster Lie algebra explicitly:
In this case
(using the notation of Section 2.3), $K = {\cal I}$, $S =\{(-1,1)\}$,
$J = \{(i,k)| i \in \Bbb Z_+, 1 \leq k \leq c(i)\}$ and $B =(2)$. Thus
$\frak g(B) =\frak g(B)^e = \frak s \frak l_2$, so that $\frak d_B=0$
and $\dim \frak h^e =1$.
As before, we write $h_{-1}$ for $h_{-1,1}$. Since for $i \in  \Bbb Z_+$
$\mu_{(i,k)}$ depends only on $i$, we may denote it $\mu_i$. Then
$\mu_i(h_{-1})= -(i-1)$ and $(\mu_i,\mu_j) = (i-1)(j-1)/2$. Hence the
entries in the $(i,j)$-block of $Q$ have value
$(i+1)(j+1)/2$. We may
therefore take $\frak z$ to be one-dimensional. If $z$ spans $\frak z$
and satisfies $(z,z)_{\frak z}=1$, then we have $\gamma_i(z) =
(i+1)/\sqrt{2}$. Note that $\frak a_0 = \frak s\frak l_2 \oplus \frak z
\cong \frak g\frak l_2$. Because the center $\frak c$ is equal to
$\mbox{Rad}(\cdot , \cdot)$, we conclude that $\frak m = \frak a_{max}$
for $\frak a = {\scr V}' \oplus \frak g \frak l_2 \oplus {\scr V}$.

\section{Generalized Verma modules and standard modules}

Given a generalized Kac-Moody Lie algebra, we observe that certain
standard modules are equal to induced modules. For Lie algebras $\frak g$
satisfying the conditions of Theorem~\ref{thm:free}, this means that
the tensor algebra over a certain vector space has the structure of an
irreducible
module for $\frak g$. In this way we show that $\frak m$ (and
$\frak g(A)/\frak c'$) can be
realized naturally as an explicitly prescribed $M$-covariant Lie
algebra of operators
on $T({\scr V})$ (which has the form of an induced $\frak m$- or
$\frak g(A)/\frak c'$-module for
certain weights $\lambda$) for our $M$-module ${\scr V}$.

\subsection{Examples of standard modules}

The conditions in the definition below of standard module appear in
\cite{B1}; the definition was made in \cite{Jur1}. Standard modules
are the only modules for which a character formula
is proven (\cite{B1}; see \cite{HMY}, \cite{Jur1} and \cite{K2}). Note that
standard modules are not the same as ``integrable highest
weight modules'' for generalized Kac-Moody algebras (see \cite{K2} for
the definition of ``integrable'' in the context of generalized Kac-Moody
algebras).

Let $X$ be a $\frak g^e$-module. A weight
$\lambda \in (\frak h^e)^*$ is {\it dominant} if $(\lambda
,\alpha_i)\in \Bbb R$ and $(\lambda ,\alpha_i) \geq 0$
for all $i \in I$. A weight $\lambda$ is called {\em integral} if
${2 \lambda(h_i) \over a_{ii}} \in {\Bbb Z}$ for all $i \in I$ such
that $a_{ii} >0$. Denote by $P_+$ the set
of dominant integral weights.

As in \cite{Jur1}, define $X$ to be
a {\it standard} module if $X$ is a highest weight module (\cite{B1}
and \cite{Jur1} actually use lowest weight modules) with highest
weight $\mu \in P_+$ and highest weight vector $x$ such that:
\begin{enumerate}
 \item  for $i \in I$, if $(\mu , \alpha_i) =0$ then
$f_i \cdot x =0$;
 \item  if $\alpha_i$ ($i \in I$) is real then
$f_i^{n_i +1} \cdot x =0$, where $n_i = 2(\mu, \alpha_i)
 /(\alpha_i,\alpha_i)$ (necessarily a nonnegative integer).
\end{enumerate}

\begin{proposition}
All standard modules are irreducible.
\end{proposition}

\noindent{Proof}: This follows from the character formula for a
standard module stated in
\cite{B1} and \cite{K2}. For two different complete proofs of the
character formula see
\cite{HMY} and \cite{Jur1}.
$\square$ \vspace{1ex}

Let $\lambda \in P_+$, and consider the standard
(irreducible) highest weight $\frak r^e$-module $L(\lambda)$ associated
to $\lambda$; it is irreducible as a
$\frak g_S^e$-module
(cf. \cite{K2}).
The highest weight space of
$L(\lambda)$ (as a $\frak g_S^e$-module) is a weight space for
$\frak h^e$, with weight $\lambda$.
Let $\frak u^+$ act trivially on $L(\lambda)$; this gives $L(\lambda)$
the structure of an irreducible $\frak p^e$-module.
Define the {\em generalized
Verma module} $V^{L(\lambda)}$ (cf. \cite{L2})
to be the induced module ${U}(\frak g^e)
\otimes_{{U}(\frak p^e)} L(\lambda)$. It is clear that $V^{L(\lambda)}
\cong {U}(\frak u^-) \otimes L(\lambda)$ as vector spaces.

We make the following observation:

\begin{proposition}
Let $\lambda \in P_+$ be an element satisfying
$(\lambda, \alpha_i) > 0$ for every imaginary simple root $\alpha_i$
 ($i  \in I\backslash S$).
Then the generalized Verma module
$V^{L(\lambda)}$ is a standard, and therefore irreducible, module for
$\frak g(A)^e$.
\label{thm:standard}
\end{proposition}

\noindent Proof: The module $V^{L(\lambda)}$ is a highest weight module with
highest weight $\lambda$, and clearly satisfies conditions $1$ and $2$
above.
$\square$ \vspace{1ex}

Proposition~\ref{thm:standard} immediately gives a class of examples of
standard modules for arbitrary generalized Kac-Moody algebras. In
the case of generalized Kac-Moody algebras satisfying the hypothesis
of Theorem~\ref{thm:free}, like $\frak m$, these generalized Verma
modules are
particularly easy to describe. As another example, we show that the
irreducible module
constructed in \cite{GT} for a generalized Kac-Moody Lie algebra with
one imaginary simple root is an example of such a generalized Verma
module.

\begin{theorem}
Let $\frak g(A) = \frak g$ be a generalized Kac-Moody algebra such that if
$\alpha_i$ and $\alpha_j$ are two distinct imaginary simple roots then
$(\alpha_i, \alpha_j)=a_{ij}<0$. Let $W$ denote the direct sum of the
integrable highest weight $\frak g_S^e$-modules ${U}(\frak n^-_S)\cdot
f_j$ for all $j \in I\backslash S$. Given $\lambda\in P_+$ such that
$(\lambda, \alpha_i) > 0$ for every
imaginary simple root $\alpha_i$ ($i \in I\backslash S$),
the generalized Verma module $V^{L(\lambda)}$ has the form
$T(W)\otimes L(\lambda)$, and is a standard (and
in particular irreducible) $\frak g^e$-module.
\label{thm:verma}
\end{theorem}

\noindent{Proof}: Let $\lambda
\in (\frak h^e)^*$ be such that $(\lambda, \alpha_i)=
\lambda(h_i)\neq 0 \mbox{ for all }i \in S$.
The generalized Verma module $V^{L(\lambda)} =
{U}(\frak u^-)\otimes {L(\lambda)}$ is a standard module by
Proposition~\ref{thm:standard}.
Since $\frak u^-$ is the free Lie algebra on $W = \coprod_{k}{U}
(\frak n^-_S) \cdot f_k$ (by Theorem~\ref{thm:free}), we conclude
that ${U}(\frak u^-) \otimes {L(\lambda)} \cong T(W)\otimes
{L(\lambda)}$.
$\square$

\begin{corollary}
In the setting of Theorem~\ref{thm:verma} assume that
$\lambda(h_i)=0
\mbox{ for all }i \in S$. Then tensor algebra $T(W)$
carries the natural structure of a
standard $\frak g^e$-module. \label{cor:verma}
\end{corollary}

\noindent Proof:
The $\frak g_S$-module
$L(\lambda)$ is one-dimensional.
$\square$ \vspace{1ex}

Note that Corollary~\ref{cor:verma} includes the case in \cite{GT}, where
$\frak g_S$ is a Kac-Moody algebra and the
generalized Kac-Moody algebra $\frak g$ is formed by adjoining one
additional imaginary simple root $\alpha_0$ to $\frak g_S$. This is
done (see \cite{GT}) by taking $\frak g$ to be the Lie algebra
associated to the matrix:
$$\l(\begin{array}{c|c}
      \l(a_{ij}\r)_{i,j \in S}  & \l(a_{0j}\r)_{ j \in S}\\
       \hline
        \l(a_{j0}\r)_{j \in S}          &  a_{00}
\end{array}\r)$$ where  $a_{00} \leq 0$.
The hypothesis of
Theorem~\ref{thm:verma} is then satisfied, because the condition on the
imaginary simple roots becomes vacuous. Let $f$ be the generator in
$\frak n^-$ associated to the one imaginary simple root.
Thus if $W$ denotes the integrable
 highest weight module
${U}(\frak n^-_S) \cdot f$, then $T(W)$ is a standard module. It
is isomorphic to the generalized Verma module associated to the module
${\Bbb C}_\lambda$ where $(\lambda, \alpha_0) = 1$ and $(\lambda,
\alpha)=0$ for simple $\alpha \in \Delta_R$.

Recall the matrix $A$ given in Section 2.3, so that the Lie algebra
$\frak g(A)$ has only only one real simple root and all of the
$a_{ij} = (\alpha_i, \alpha_j) <0$ for $\alpha_i, \alpha_j$
imaginary simple, so the
conditions of Corollary~\ref{cor:verma} are satisfied.
In the case of the Lie algebra $\frak g(A)/\frak c'$ of Proposition
\ref{prop:c} and the Monster
Lie algebra $\frak m$, we can obtain standard modules by applying
Corollary~\ref{cor:verma} to $\frak g(A)$.
If we choose $\lambda \in P_+$ satisfying the conditions
\begin{equation}
\begin{array}{lllc}
\lambda(h_{-1})& =& 0 &  \\
\lambda(h_{ik})& =& \lambda(h_{ij}) & \mbox{ for each  } i>0,\  1 \leq
k,j \leq c(i),
\end{array} \label{eq:mc'}
\end{equation}
we get an irreducible generalized Verma module $V^{L(\lambda)}$ for
$\frak g(A)/ \frak c'$.
To obtain $\frak m$-modules, given any real numbers $a, b$ such that $a>0$
and $a+b >0$ let
$\lambda \in \frak (h^e)^*$ be an element satisfying the conditions
\begin{equation}
\begin{array}{lllc}
\lambda(h_{-1})& =& 0 &  \\
\lambda(h_{ik})& =& \lambda(h_{ij})= ai +b  & \mbox{ for each  } i>0,
\ 1 \leq k,j \leq c(i).
\end{array} \label{eq:mc}
\end{equation}
Then $\lambda \in P_+$, and the full center $\frak c$ of $\frak g(A)$
acts as zero
on the module $V^{L(\lambda)}$. Thus
$V^{L(\lambda)}$ is also an $\frak m$-module.

Recall the $M$-module ${\scr V}$ given by (\ref{eq:ik}), so that
$\frak u^-= L( {\scr V} )$.
We have shown that for the choices of $\lambda$ given above the
$M$-module $T({\scr V})=U(\frak u^-)$ has a natural standard module structure
for $\frak g(A)/\frak c'$ or for $\frak m$.

Summarizing the above considerations, we describe $\frak m =L({\scr V})
\oplus \frak g \frak l_2 \oplus L({\scr V}')$ (and $\frak g(A)/\frak c'$) in
the following (obvious) way:

\begin{theorem}
Let ${\scr V}$ be given by (\ref{eq:ik}).
The Lie algebra $\frak m$ can be realized in a natural way as a Lie
algebra of operators on the
irreducible $\frak m$-module $T({\scr V})$, identified as above with a
generalized Verma module, for any weight $\lambda$
satisfying (\ref{eq:mc}). The $M$-module structure on the tensor
algebra $T({\scr V})$ induced by the $M$-module structure on ${\scr V}$ is
compatible with the action of $M$ on $\frak m$.
The following operators
generate $\frak m$: ${\scr V}$ acting on $T({\scr V})$
by left multiplication, $\frak g
\frak l_2$ acting on $T({\scr V})$ via the action induced by the $\frak g
\frak l_2$-module structure of ${\scr V}$, and ${\scr V}'$ acting on
$T({\scr V})$ via its
natural action on the induced module.
Furthermore, these comments extend
to the Lie algebra $\frak g(A)/\frak c'$ for weights $\lambda$ satisfying
(\ref{eq:mc'}). $\square$
\end{theorem}

We suggest that these irreducible $\frak m$-modules $T(\scr V)$ (and
similarly, $T(\scr V')$) are realizable in a conceptual way as the
particular free associative algebras of operators described in our
conjecture in Section 2.

\section{Application to Borcherds' proof of the moonshine conjectures}

We will apply the decomposition of $\frak m$ as $\frak u^+ \oplus
\frak g \frak l_2 \oplus \frak u^-$, where $\frak u^+$ and $\frak u^-$
are free Lie algebras, to
simplify part of the proof appearing in \cite{B3} of the
Conway-Norton conjectures \cite{CN}. It has already been seen
\cite{Jur2} that the
denominator identity for $\frak m$ can be obtained with the help of
the
fact that $\frak u^-$ is a free Lie algebra (but still using Borcherds'
product formula for $j(q)$). This provides one
simplification of Borcherds' proof, because it requires less of
the theory of Kac-Moody algebras to be generalized.
Another simplification will be
obtained here by computing the homology $H(\frak u^-,
{\Bbb C})$ of the subalgebra $\frak u^-$,
rather than computing the
homology of the Lie algebra $\frak n^+$ (or equivalently $\frak n^-$)
as in \cite{B3}. This avoids using results of \cite{GL}
extended to generalized Kac-Moody Lie algebras.

We show that the replication formulas of \cite{CN} can be obtained
{}from this homology result and calculations similar to those appearing
in \cite{Bou} for computing the dimensions of homogeneous subspaces of
free Lie algebras.

\subsection{Homology}

It is easy
to compute the homology of the free Lie algebra $L(V)$ for a vector
space $V$:
The following exact sequence is a ${U}(L(V)) = T(V)$-free resolution
of the trivial module:
\begin{equation}
0 \rightarrow T(V) \otimes V {\buildrel \mu  \over \rightarrow} T(V)
{\buildrel \epsilon  \over \rightarrow} {\Bbb C}  \rightarrow 0
\label{eq:res}
\end{equation}
where $\mu$ is the multiplication map and $\epsilon$ is the
augmentation map. Note that if a group or a Lie
algebra acts on $V$
then the group or Lie algebra also acts on every term in the
resolution (\ref{eq:res}) in a
natural way, and therefore acts on $H( L(V), \Bbb C)$.

One immediately observes from the resolution (\ref{eq:res}):
$$
H_0(L(V),{\Bbb C}) = \Bbb C \label{eq:h0}
$$
\begin{equation}
H_1(L(V),{\Bbb C}) =  V \cong L(V)/ [L(V), L(V)]
\label{eq:h1}
\end{equation}
$$
H_n( L(V) ,{\Bbb C}) =0 \mbox{ for } n \geq 2.
$$
(It is also well known how to compute $H_0(\frak a,\Bbb C)$ and
$H_1(\frak a,\Bbb C)$ for an arbitrary Lie algebra $\frak a$
\cite{CE}.)

\subsection{Adams operations}

In this section $p$, $q$ and $t$ are commuting formal variables. The
variables
$p^{-1}$ and $q^{-1}$ will be used to
index the $\Bbb Z \oplus \Bbb Z$-grading of our vector spaces.
All of the $M$-modules we encounter are finite-dimensionally
$\Bbb Z \oplus \Bbb Z$-graded with grading suitably truncated and will
be identified with formal series in $R(M)[[p,q]]$. Definitions and
results from \cite{Knu} about the
$\lambda$-ring $R(M)$ of finite-dimensional representations of $M$ are
applicable to formal series in $R(M)[[p,q]]$. We summarize the results
of \cite{At}, \cite{Knu} that we use below.

Let $G$ be any finite group. The
representation ring $R(G)$ is a $\lambda$-ring \cite{Knu} with the
$\lambda$ operation given by exterior powers, so $\lambda^i V =
\bigwedge^i V$ for $V \in R(G)$.

In the following discussion we let $W,V \in R(G)$. The operation
$\bigwedge^i$ satisfies
\begin{equation}
{\textstyle\bigwedge^i} (W\oplus V)
=\sum_{n=0}^i {\textstyle\bigwedge^n(W)\otimes \bigwedge^{i-n}}(V) .
\label{eq:rul}
\end{equation}
Define
\begin{equation}\textstyle
\bigwedge_t(W) = \bigwedge^0 (W) +\bigwedge^1(W)t +\bigwedge^2 (W)t^2 +\cdots.
\label{eq:wedge}
\end{equation}
Then
\begin{equation}\textstyle
\bigwedge_t (V \oplus W) = \bigwedge_t(V) \cdot \bigwedge_t(W).
\label{eq:times}
\end{equation}
The Adams operations
 $\Psi^k : R(G) \rightarrow R(G)$ are defined for
$W \in R(G)$ by:
\begin{equation}
{\textstyle {d \over dt} \log \bigwedge_t(W)} = \sum_{n \geq 0} (-1)^n
\Psi^{n+1} (W) t^n .
\label{eq:log}
\end{equation}
We use the following properties of the $\Psi^k$
{}from \cite{Knu}:
\begin{equation}
\Psi^k (V\otimes W) = \Psi^k (V)\otimes \Psi^k (W) \label{eq:psi1}
\end{equation}
\begin{equation}
\Psi^k (V\oplus W) = \Psi^k (V)\oplus\Psi^k (W) .
\end{equation}
Thus the $\Psi^k$ are ring homomorphisms. Also
\begin{equation}
\Psi^k\Psi^l (W) = \Psi^{kl}(W) .\label{eq:psi2}
\end{equation}
For a class function $f : G \rightarrow \Bbb C$, define
\begin{equation}
(\Psi^k f)(g) = f(g^k).
\end{equation}
for all $g \in G$.
Then if ${\chi}_V$ is a character of $G$,
\begin{equation}
\chi_{\Psi^k(V)}(g)= (\Psi^k \chi_V )(g)
\end{equation}
(cf. \cite{Knu}).

Now let $W$ be a finite-dimensionally $\Bbb Z \oplus \Bbb Z$-graded
representation of $G$ such that $W_{(\gamma_1, \gamma_2)}=0$ for
$\gamma_1, \gamma_2 >0$. We shall write
\begin{equation}
W =\sum_{(\gamma_1,\gamma_2) \in \Bbb N^2}W_{(-\gamma_1,
-\gamma_2) }p^{\gamma_1}q^{\gamma_2},\label{eq:fgra}
\end{equation}
identifying the graded space and formal series.
We extend the definition of $\Psi^k$ to formal series $W\in R(G)[[p,q]]$
by defining $\Psi^k (p)= p^k$, $\Psi^k (q)= q^k$ and in general,
\begin{equation}
\Psi^k ( \sum_{(\gamma_1,\gamma_2) \in \Bbb N^2}W_{(-\gamma_1,
-\gamma_2)}p^{\gamma_1}q^{\gamma_2}) =
        \sum_{(\gamma_1,\gamma_2) \in \Bbb N^2}\Psi^k
(W_{(-\gamma_1, -\gamma_2)}) p^{k\gamma_1}q^{k\gamma_2}.
\label{eq:def}\end{equation}

Expressions involving $\bigwedge^i (W)$ for $W \in R(G)[[p,q]]$ are
interpreted in the obvious way, so that (\ref{eq:wedge}) is
meaningful and
(\ref{eq:rul}) and (\ref{eq:times}) are valid for our
finite-dimensionally graded spaces.
The identity (\ref{eq:log}) remains valid for $W\in R(G)[[p,q]]$ and
the $\Psi^k$ defined above by (\ref{eq:def}). Also, (\ref{eq:psi1}) -
(\ref{eq:psi2}) remain valid.

\subsection{M\"{o}bius inversion}
Let $G$ be a finite group and let $W$ be a finite-dimensionally
$\Gamma$-graded representation of $G$, where $\Gamma$ is a finitely
generated free abelian group. Then $W =\coprod_{\gamma \in \Gamma}
W_\gamma$ where
each $W_\gamma$ is a finite-dimensional module for $G$.
We prove a M\"{o}bius inversion formula involving
functions $s,t: G \times \Gamma \rightarrow \Bbb C$ such that for
fixed $\gamma \in \Gamma$, $s$ and $t$ are class functions of $G$. For
each $g \in G$, $s$ and $t$ are functions taking $\Gamma$ to $\Bbb C$;
we sometimes write $s(\gamma)$, $t(\gamma)$, $\gamma \in
\Gamma$. For example, we
may take
$t(g, \gamma) = Tr(g|W_\gamma)$, $g \in G,\  \gamma \in \Gamma$.
If $\Gamma$ has base
$\{\gamma_1, \ldots, \gamma_k\}$, and if $\nu = a_1\gamma_1 + \cdots +
a_k\gamma_k \in \Gamma$ ($a_i \in \Bbb Z$) and $d \in \Bbb Z$ then we
say $d | \nu$ if $d |a_i$ for each $i$. For $\nu, \kappa \in \Gamma$
such that $ \nu = d \kappa$ we write ${\nu \over \kappa} =d$.
Let $\mu$ be the M\"{o}bius function.
\begin{lemma}
Let $s$ and $t$ be as above and let $\nu \in \Gamma$, $d \in \Bbb Z_+$.
Then
\begin{equation}
\sum_{d|\nu} \Psi^d t\l({\nu\over d}\r) = s(\nu)
\label{eq:mo1}
\end{equation}
if and only if
\begin{equation}
t(\nu) = \sum_{d|\nu} \mu(d)\Psi^d s\l({\nu\over d}\r).
\label{eq:mo2}
\end{equation}
\end{lemma}

\noindent Proof: We use a standard argument (cf. \cite{Bou}).
Assume equation (\ref{eq:mo1}). Then
$$ \sum_{d|\nu} \mu(d)\Psi^d s\l({\nu\over d}\r) =  \sum_{d|\nu} \mu(d)\Psi^d
\l(\sum_{\delta|{\nu \over d}} \Psi^{\delta} t\l({\nu\over d \delta}\r)\r)$$
\begin{eqnarray*}
& =& \sum_{d|\nu} \mu(d) \sum_{\delta|{\nu \over d}} \Psi^{d\delta}
 t\l({\nu\over d \delta}\r)\\
& =& \sum_{d \delta |\nu}  \mu(d) \Psi^{d\delta}  t\l({\nu\over d \delta}\r)\\
& =& \sum_{\kappa|\nu}  \sum_{d |{\nu \over \kappa}} \mu(d)\Psi^{\nu/\kappa}
t(\kappa)\\
& =& \sum_{\kappa|\nu} \Psi^{\nu/\kappa} t(\kappa)   \sum_{d |{\nu \over
\kappa}} \mu(d) = t(\nu).
\end{eqnarray*}
That (\ref{eq:mo2}) implies (\ref{eq:mo1}) is similar.
$\square$ \vspace{1ex}

We will apply this inversion formula to the situation where $\Gamma =
\Bbb Z \oplus \Bbb Z$ and $W \in R(G)[[p,q]]$ as in (\ref{eq:fgra}).
For a pair $(i,j)
\in \Bbb Z \oplus \Bbb Z$ our definition specializes to:
$k|(i,j)$ if $k(m,n)=(i,j)$ for some $(m,n) \in \Bbb Z\oplus \Bbb Z$.
For $(i,j) \in \Bbb Z_+ \oplus \Bbb Z_+$ we define
$$P(i,j) = \{ a=(a_{rs})_{r,s \in \Bbb Z_+}\ |\
a_{rs}\in \Bbb N,\ \sum_{(r,s)\in \Bbb Z_+\oplus \Bbb Z_+}
a_{rs}(r,s)=(i,j)\}.$$
We will use the notation $|a| = \sum a_{rs}$, $a! = \prod a_{rs}!$.

\subsection{Replication}

In this section we show that the formulas for the graded traces of
elements of $M$ acting on $V^\natural$ (i.e., the McKay-Thompson
series) are replicable \cite{CN}, in
fact, we show that they are completely replicable in
the sense of \cite{No} (see also \cite{CuN} and \cite{F}). We also observe
that they can be computed recursively, as in \cite{B3}. The free Lie
algebra structure of
$\frak u^-$ allows us to perform a calculation analogous to one
appearing in \cite{Bou}, computing the dimensions of homogeneous
subspaces of a free Lie algebra. In this context the replication
formulas occur quite naturally.

Recall the structure of $\frak u^-$ and ${\scr V} =
H_1(\frak u^-)$ as
$\Bbb Z \oplus \Bbb Z$-graded $M$-modules ((\ref{eq:um}),
Lemma \ref{lem:ng}, Lemma \ref{lem:V}, (\ref{eq:h1})). We index the
grading by
$p^{-1}$ and $q^{-1}$ as in (\ref{eq:fgra}); then write $\frak u^-$
and ${\scr V} = H_1(\frak u^-)$ as elements of $R[M][[p,q]]$:
\begin{equation}
\frak u^- = \sum_{(m,n)} V^\natural_{mn} p^m q^n \label{eq:u}
\end{equation}
and
\begin{equation}
{\scr V} = \sum_{(m,n)} V^\natural_{m+n-1} p^m q^n, \label{eq:v}
\end{equation}
where here and below the sums are over all pairs $(m,n)$ such that $m,n >0$.

Define
$$H_t(\frak u^-) = \sum_{i=0}^\infty H_i (\frak u^-) t^i$$
and let $H(\frak u^-)$ denote the alternating sum $H_t(\frak
u^-)|_{t={-1}}$.
Recall the Euler-Poincar{\'e} identity:
\begin{equation}\textstyle
\bigwedge_{-1}(\frak u^-) = H(\frak u^-). \label{eq:hom}
\end{equation}
Taking $\log$ of both sides of (\ref{eq:hom}) results in the formal power
series identity in $R(M)[[p,q]]\otimes \Bbb Q$:
\begin{equation}
\textstyle
 \log \bigwedge_{-1}(\frak u^-) =  \log H (\frak u^-),\label{eq:lg}
\end{equation}
where we have
$$\log H (\frak u^-)= \log(1 - H_1(\frak u^-)) = -\sum_{n =1}^\infty
{1 \over n}H_1(\frak u^-)^n. $$
Formally integrating (\ref{eq:log}), with $W = \frak u^-$, gives
$$
{\textstyle \log \bigwedge_{t}}(\frak u^-)=
-\sum_{n \geq 0}  \Psi^{n+1} (\frak u^-) {(-t)^{n+1}\over n+1}.
$$
Then setting $t =-1$ gives:
$$ -\log {\textstyle\bigwedge_{-1}}(\frak u^-) =  \sum_{k=1}^\infty
{1 \over k}\Psi^k(\frak u^-).$$
Since $H_1(\frak u^-) = {\scr V}$, equation (\ref{eq:lg}) and
equations (\ref{eq:u}) and (\ref{eq:v}) imply
$$\sum_{k=1}^\infty {1 \over k}\Psi^k(\sum_{(m,n)} V^\natural_{mn}p^m
q^n)
= \sum_{k=1}^\infty {1 \over k}( \sum_{(m,n)} V^\natural_{m+n-1}p^m
q^n)^k$$

$$\sum_{k=1}^\infty \sum_{(m,n)}{1 \over k}\Psi^k( V^\natural_{mn})p^{mk}
q^{nk}
=  \sum_{k=1}^\infty {1 \over k}\sum_{(i,j)\in \Bbb Z_+}\sum_{a \in
P(i,j)\atop |a|= k}{|a|!\over a!} \prod_{r,s \in \Bbb Z_+}
(V^\natural_{r+s-1})^{a_{rs}} p^i q^j
$$

$$\sum_{(i,j)} \sum_{k|(i,j)}{1 \over k}\Psi^k(
V^\natural_{ij/k^2}) p^{i}q^{j}
=  \sum_{(i,j)} \sum_{a \in P(i,j)}{(|a|-1)!\over
a!} \prod_{r,s \in \Bbb Z_+} (V^\natural_{r+s-1})^{a_{rs}} p^i q^j.$$

Take the trace of an element $g \in M$ on both sides of this formula,
writing $c_g(m)$ for $\mbox{Tr}(g|V_m)$:

$$\sum_{(i,j)} \sum_{k| (i,j)}{1\over k}\Psi^k(
c_g(ij/k^2)) p^{i}q^{j}
=  \sum_{(i,j)}\sum_{a \in P(i,j)}{(|a|-1)!\over a!} \prod_{r,s \in \Bbb Z_+}
c_g(r+s-1)^{a_{rs}} p^i q^j. $$
Equating the coefficients of $p^iq^j$ gives
\begin{equation}
H_{i,j} =  \sum_{k| (i,j)}{1\over k}H^{(k)}_{ij/k^2}, \label{eq:no}
\end{equation}
where, using the notation of \cite{No}, we set (for each $g \in M$)
$$H_i^{(k)} = c_{g^k}(i)$$
and
$$H_{i,j} = \sum_{a \in P(i,j)}{(|a|-1)!\over a!} \prod_{r,s \in \Bbb Z_+}
c_g(r+s-1)^{a_{rs}}.$$
Formula (\ref{eq:no}) asserts precisely that the McKay-Thompson series
$$T_g(V^\natural) =\sum_{i \in \Bbb Z}c_g(i)p^i, $$
$g \in M$, satisfy the
``replication formula'' as it is written in \cite{No} (see also
\cite{CN}, \cite{CuN} and \cite{F}).

 Recall the definition of ``replicability'' given in \cite{No}: that
$H_{i,j} = H_{m,n}$ whenever $ij = mn$ and
gcd$(i,j)=\,$gcd$(m,n)$. (See also \cite{CuN}, where this definition is
related to the replication formulas of \cite{CN}.)
Since all of
the ``replicates'' $T_{g^k}(V^\natural)$ of the McKay-Thompson series
$T_{g}(V^\natural)$ are replicable, the $T_{g}(V^\natural)$ are
``completely replicable'' (\cite{No}; see also \cite{F}). Thus we observe
{}from (\ref{eq:no}):

\begin{proposition}
The McKay-Thompson series $T_g(V^\natural) =\sum_{i \in \Bbb
Z}c_g(i)p^i $, $g \in M$, are completely replicable functions. $\square$
\end{proposition}
\vspace{1ex}

The modular functions of \cite{CN} also satisfy
(\ref{eq:no}),
by \cite{No}, \cite{Ko}, \cite{CuN} and \cite{F}.

In order to obtain recursion
relations for the McKay-Thompson series, we apply M\"{o}bius inversion
to (\ref{eq:no}), with the functions
$t((i,j)) = (i+j)c_g(ij)$ and $s((i,j))= (i+j)H_{i,j}$:
$$c_g(ij) = \sum_{k>0 \atop k(m,n) = (i,j)}{1\over k}\mu(k) \Psi^k
\l(\sum_{a
\in P(m,n)} {(|a|-1)!\over a!} \prod_{r,s \in \Bbb Z_+}
c_g(r+s-1)^{a_{rs}}\r)$$
\begin{equation}
= \sum_{k>0 \atop k(m,n)= (i,j)}{1\over k} \mu(k)\l(\sum_{a
\in P(m,n)} {(|a|-1)!\over a!} \prod_{r,s \in \Bbb Z_+}
c_{g^k}(r+s-1)^{a_{rs}}\r). \label{eq:rep}\end{equation}

If $Q$ is a polynomial expression in the $c_g(n)$, $g\in M$, $n \leq
N$ we say that $Q$ has {\em level} $\leq N$. Then (\ref{eq:rep}) implies
that
\begin{equation}
c_g(ij) = c_g(i+j-1) + \mbox{ an expression of level }\leq i+j-3.
\label{eq:ep}
\end{equation}
This observation is, of course, vacuous if $i$ or $j =1$. Taking
$(i,j) = (2l,2)$ equation (\ref{eq:ep}) gives
$$c_g(4l) = c_g(2l+1) + \mbox{ an expression of level }\leq 2l-1.$$
Taking $(i,j) =(2l+1,2)$ equation (\ref{eq:ep}) gives
$$c_g(4l +2) = c_g(2l+2) + \mbox{ an expression of level }\leq 2l.$$
For $(i,j) = (l,4)$ equation (\ref{eq:ep}) gives
$$c_g(4l)= c_g(l+3) + \mbox{ an expression of level }\leq l+1.$$
It follows that the value of $c_g(4l)$ is given by an expression of
lower level whenever $4l > 2l +1$, i.e., for $l \geq 1$. Also,
$c_g(4l+2)$ is given by an expression of lower level whenever $4l+2 >
2l+2$, i.e., for $l \geq 1$. Finally, equating the two expressions for
$c_g(4l)$ shows that $c_g(2l+1)$ is given by an expression of lower
level whenever $2l+1 >l+3$, i.e., for $l >2$. Thus $c_g(n)$ is
determined by expressions of lower level except when $n =1,2,3,5$.
Thus the values of the $c_g(n)$ are determined by the $c_h(1)$,
$c_h(2)$, $c_h(3)$, $c_h(5)$, $h \in M$, and equation (\ref{eq:rep}).

As in \cite{B3}, we conclude that
since both the McKay-Thompson series and the
modular functions of \cite{CN} satisfy (\ref{eq:rep}),
all that is necessary to prove that these functions
are the same (see \cite{B3}) is to check the initial data listed above.

\end{document}